\pdfoutput=1
\documentclass[pre,aps]{revtex4}

\usepackage{times}
\usepackage{amsmath}
\usepackage{graphicx}
\usepackage{amssymb}
\usepackage{xcolor}

\begin{document}

\title{Vortex Pairs in the Discrete Nonlinear Schr{\"o}dinger Equation}

\author{J. J. Bramburger}
\affiliation{Division of Applied Mathematics, Brown University, Providence, RI 02906, USA}

\author{J. Cuevas-Maraver}
\affiliation{Grupo de F\'{i}sica No Lineal, Departamento de F\'{i}sica Aplicada I,
Universidad de Sevilla. Escuela Polit\'{e}cnica Superior, C/ Virgen de \'{A}frica, 7, 41011-Sevilla, Spain \\
Instituto de Matem\'{a}ticas de la Universidad de Sevilla (IMUS). Edificio Celestino Mutis. Avda. Reina Mercedes s/n, 41012-Sevilla, Spain}

\author{P. G. Kevrekidis}
\affiliation{Department of Mathematics and Statistics, University of Massachusetts
Amherst, Amherst, MA 01003-4515, USA}

\begin{abstract}
The aim of the present work is to examine the role of discreteness in the interaction of both co-winding and counter-winding vortices in the context of the nonlinear Schr{\"o}dinger equation. Contrary to the well-known rotation of same charge vortices, and translation of opposite charge vortices, we find that strong discreteness is able to halt both types of pairs into  stationary, potentially stable configurations up to a critical inter-site coupling strength. Past the relevant critical point the behavior is also somewhat counter-intuitive as, for instance, counterwinding vortices start moving but also approach each other. This lateral motion becomes weaker as the continuum limit is approached and we conjecture that genuine traveling appears only at the continuum limit. Analogous features arise in the cowinding where the discrete coherent structure pair spirals outward, with rigid rotation being restored only in the continuum limit.
\end{abstract}

\maketitle

\section{Introduction} 

The discrete nonlinear Schr{\"o}dinger equation (DNLS) constitutes one of the
most prototypical examples of a nonlinear dynamical lattice, combining the linear
form of lattice (discrete) dispersion and nonlinearity~\cite{dnlsbook}.
For this reason the model has been argued to be relevant as an exact
or asymptotic description of a variety of different settings including,
but not limited to, optical waveguide arrays~\cite{dnc,moti},
as well as the evolution of atomic Bose-Einstein condensates (BECs) in the
presence of optical lattice potentials~\cite{ober}. These applications
have been motivated by the theoretical exploration and even experimental observation
of a diverse host of features such as discrete diffraction~\cite{yaron} and diffraction management~\cite{yaron1},
lattice solitary waves~\cite{yaron,yaron2} and discrete vortices~\cite{neshev,fleischer},
Talbot revivals~\cite{christo2}, and $\mathcal{PT}$-symmetry
breaking~\cite{kip}, among many others.

Especially in two-dimensional settings, the study of both waveguide
arrays and also photorefractive crystals has offered a wide range
of possibilities~\cite{moti}. Most recently, this includes, e.g.,
the study of
topologically protected states in variants of the lattices that
break the time-reversal symmetry~\cite{moti2,leykam,mark2}. However,
many of the relevant studies have been conducted in the focusing
nonlinearity realm where bright solitonic structures on top of
a vanishing background may exist. While gap structures have been
considered in the defocusing realm in square~\cite{hadii} and non-square lattices~\cite{law},
there is considerably less effort in the subject of vortices and their
associated dynamics.

Indeed, vortex dynamics and interactions are of principal relevance to the
evolution of atomic Bose-Einstein condensates~\cite{fetter1,fetter2,siambook}. Furthermore,
BECs often involve the evolution in periodic potentials~\cite{ober}, which
 in recent two-dimensional extensions have even been considered in the realm
 of geometries with curvature~\cite{porto}.
Nevertheless, the concurrent exploration of defocusing nonlinearity-induced
vortices and discreteness has been quite limited, to the best of our knowledge,
and in fact has been constrained to the study of a single such entity~\cite{1Vortex,Bramburger}.
The aim of the present work is to go a step past this and develop a systematic
understanding of the principal numerical phenomenology, aided by some analytical
insights, of the case of multiple vortices in the DNLS model.
This is a topic of interest for a number of reasons: continuum vortex pairs have
a very definite behavior dictated by the topological charges.
For same-charge (cowinding) vortices, the result of their interaction is a rigid
rotation around their center of mass, while for opposite-charge (counterwinding)
vortices, the coherent structures move parallel to each other in a steady
translational (constant speed) motion~\cite{fetter1,fetter2,siambook}. Discreteness,
on the other hand, is well-known to ``disrupt'' the translational dynamics of solitary waves,
due to the so-called Peierls-Nabarro barrier~\cite{dnlsbook}. Hence, it appears to
be of particular interest what the result of the interplay of these opposing tendencies is.

Our findings can be summarized in the following conclusions:
\begin{itemize}
    \item For sufficiently weak coupling, discreteness ``dominates'' the interaction,
    entirely halting the rotational or translational motion of counter- or co-winding
    vortices, and leading instead to the formation of stable stationary configurations
    of such states.
    \item Past a sufficiently large critical coupling, the relevant branches feature
    a turning point bifurcation and stationary states cease existing.
    A one-dimensional example of such a saddle-center bifurcation has appeared
    for dark solitons in the work of~\cite{susjoh}.
    Interestingly,
    this saddle-center bifurcation is not the only one taking place in the system;
    there is also a pitchfork bifurcation occurring near the turning point with an
    asymmetric (or 1 vortex, as we call it) branch.
    \item Past the turning point, a reasonable expectation might be
    that traveling arises, e.g., via a SNIPER bifurcation as happens in a different context
    in discrete systems~\cite{sniper_yannis}. Nevertheless, to our surprise, we find
    that this is {\it not} the case. Instead, {\it no} traveling (for counter-winding)
    or rotating (for cowinding vortices) state exists in the dynamics past the critical
    point. Instead, cowinding vortices increase their separation distance slowly,
    while counterwinding ones move closer to each other and may eventually participate
    in catastrophic (annihilation) collisional events.
    \item As the continuum limit is approached, these ``lateral'' motions become slower,
    leading us to conjecture that genuine rotational (for cowinding) and translational (for counterwinding)
    vortex configurations can be reached solely in the singular continuum limit of the model.
\end{itemize}

The structure of our presentation is as follows. We first provide the general mathematical
formulation of the model of interest. We then simultaneously consider both the counterwinding and cowinding
cases in section III. A connection with the continuum
limit is offered in section IV. Finally, section V summarizes our findings and presents
some directions for future study.

\section{Formulation} 

Our starting point will be  a two-dimensional discrete nonlinear Schr\"odinger equation
\begin{equation}\label{dNLS}
	\mathrm{i}\frac{d\psi_{n,m}}{dt} - |\psi_{n,m}|^2\psi_{n,m} + \frac{\varepsilon}{2}\Delta\psi_{n,m} = 0, { \quad (n,m)\in\mathbb{Z}^2}
\end{equation}
where $\Delta\psi_{n,m} = \psi_{n+1,m} + \psi_{n-1,m} + \psi_{n,m+1} + \psi_{n,m-1} - 4\psi_{n,m}$ is the discrete Laplacian.
To consider potentially stationary states in the model, { for all $(n,m)\in\mathbb{Z}^2$} we introduce the ansatz
\[
	\psi_{n,m}(t) = \sqrt{\omega}\phi_{n,m}\mathrm{e}^{-\mathrm{i}\omega t},
\]
{where $\phi_{n,m}$ is time-independent,} to transform (\ref{dNLS}) to
\begin{equation}\label{dNLS2}
	\frac{C}{2}\Delta\phi_{n,m} + (1-|\phi_{n,m}|^2)\phi_{n,m} = 0, {\quad (n,m)\in\mathbb{Z}^2}.
\end{equation}
{Here we have set $C = \varepsilon/\omega$.}

Through an amplitude-phase decomposition (often referred to as the Madelung transformation~\cite{siambook}), the
complex field  is rewritten as
$\phi_{n,m} = r_{n,m}\mathrm{e}^{\mathrm{i}\theta_{n,m}}$ for all $(n,m) \in \mathbb{Z}^2$ so that solving (\ref{dNLS2}) is equivalent to solving
\begin{subequations}\label{Polar_dNLS}
	\begin{equation}\label{Polar_dNLS_Radial}
		0 = \frac{C}{2}\sum_{n',m'} (r_{n',m'}\cos(\theta_{n',m'} - \theta_{n,m}) - r_{n,m}) + r_{n,m}(1 - r_{n,m}^2), \\
	\end{equation}
	\begin{equation}\label{Polar_dNLS_Phase}
		0 = {\frac{C}{2}}\sum_{n',m'} r_{n',m'}\sin(\theta_{n',m'} - \theta_{n,m})
	\end{equation}
\end{subequations}
for each $(n,m) \in \mathbb{Z}^2$. The sum in (\ref{Polar_dNLS}) is taken over all four nearest neighbours of ($n,m$) so that $(n',m') = (n\pm 1,m),(n,m\pm 1)$.

{In this work we begin by focusing on the behaviour of solutions in the anti-continuum limit, $C \to 0^+$. Simply evaluating (\ref{Polar_dNLS}) at $C = 0$ will of course trivially solve (\ref{Polar_dNLS_Phase}), but this gives no indication as to the continuity of solutions into $C > 0$ since we no longer automatically satisfy (\ref{Polar_dNLS_Phase}) for $C \neq 0$. Therefore, to maintain continuity of solutions as $C \to 0^+$ we can replace (\ref{Polar_dNLS_Phase}) with
\begin{equation}\label{Polar_dNLS_Phase2}
	0 = \sum_{n',m'} r_{n',m'}\sin(\theta_{n',m'} - \theta_{n,m})
\end{equation}
for all $(n,m)\in\mathbb{Z}^2$ since $C/2$ appears only as a multiplicative constant in (\ref{Polar_dNLS_Phase}).}

We consider the existence and stability of vortex pair solutions of (\ref{Polar_dNLS_Radial}) and (\ref{Polar_dNLS_Phase2}) which satisfy $r_{n,m} \to 1$ when $(n,m) \to \infty$. In our work we analyze the existence, stability and dynamics of different vortex pair states in $N\times N$ lattices, with $N = 41,81,$ and $251$ (but also examine the dependence of the results
on the lattice size $N$). The vorticity of each structure is assigned to be either $S = 1$ (if the phase rotates counter-clockwise) or $S=-1$ (if it rotates clockwise) in the limit $C \to 0^+$. Either when we want to examine the unstable dynamics of the model, or when we consider values of $C$ past the critical point of existence of stationary configurations (see details below), the full dynamical model of Eq.~(\ref{dNLS}) is evolved in time. We now turn to the consideration of the two different cases of vortex pairs.

\section{Vortex Solutions} 

In this section we handle both counter- and cowinding vortex solutions together. In $\S$~\ref{subsec:Sym} we describe how the internal symmetries of the vortices can be used to reduce the number of equations required to solve (\ref{Polar_dNLS}). $\S$~\ref{subsec:Existence} presents our numerical existence and continuation results which show that both counter- and cowinding vortices as solutions of (\ref{Polar_dNLS}) can only exist up to some finite $C > 0$, after which they become dynamic solutions of the full DNLS (\ref{dNLS}). The stability of these static solutions is examined in $\S$~\ref{subsec:Stability} and then in $\S$~\ref{subsec:Dynamics} we provide dynamic simulations of the solutions near their respective critical existence thresholds in $C > 0$.

\subsection{Symmetries and Reductions}\label{subsec:Sym} 

We can obtain stationary vortex solutions by exploiting the symmetries of the system (\ref{Polar_dNLS}) and the underlying lattice structure in a similar way to what  was done for single vortex solutions in \cite{Bramburger}. Here we define a function for which we will show that its roots can be used to obtain vortex solutions of (\ref{Polar_dNLS}) with the boundary conditions giving that the vortex is either counter- or cowinding. Define
\begin{equation}\label{F_Function}
	\begin{split}
		F^1_{n,m}(C,r,\theta) &= \frac{C}{2}\sum_{n',m'} (r_{n',m'}\cos(\theta_{n',m'} - \theta_{n,m}) - r_{n,m}) + r_{n,m}(1 - r_{n,m}^2), \\
		F^2_{n,m}(C,r,\theta) &= \sum_{n',m'} r_{n',m'}\sin(\theta_{n',m'} - \theta_{n,m}),
	\end{split}
\end{equation}
for all integers $n,m\geq 0$, where $r = \{r_{n,m}\}_{n,m\geq 0}$ and $\theta = \{\theta_{n,m}\}_{n,m \geq 0}$. For some fixed $c \geq 0$, the indices $(n,m) = (\pm c, 0)$ will be considered the centers of each of the vortices. Notice that the form of $F^1_{n,m}(C,r,\theta)$ is taken to correspond to (\ref{Polar_dNLS_Radial}) and the form of $F^2_{n,m}(C,r,\theta)$ corresponds to (\ref{Polar_dNLS_Phase2}), and moreover $F^2_{n,m}(C,r,\theta)$ has no explicit dependence on $C$. Nevertheless, we will always be solving for roots of $F^1$ and $F^2$ together, endowing $F^2$ with an implicit dependence on $C$ coming from obtaining roots of $F^1$ at specific parameter values of $C$.

The system (\ref{F_Function}) is not fully defined until it is coupled with appropriate boundary conditions which account for neighboring connections with $n = -1$ or $m = -1$. It is exactly these boundary conditions that are used to extend to either counter- or cowinding vortices. We begin with counterwinding vortices and introduce the boundary conditions
\begin{equation}\label{Counter_Boundary}
	r_{-1,m} = r_{1,m}, \quad r_{n,-1} = r_{n,1}, \quad \theta_{-1,m} = \theta_{1,m}, \quad \quad \theta_{n,-1} = 2\pi - \theta_{n,1}.
\end{equation}
for all $n,m\geq 0$. Based upon these conditions, we necessarily have that $\theta_{0,m},\theta_{n,0}\in\{0,\pi\}$ for all $n,m\geq 0$. For the positive integer $c$ which is used to define the center of each of the vortices we will take
\begin{equation}\label{CountAssignment}
	\theta_{n,0} = \left\{
     		\begin{array}{cl} 0, & 0 \leq n \leq c \\ \\
		\pi, & n > c.
		\end{array}
   	\right.
\end{equation}
We note that these boundary conditions (\ref{Counter_Boundary}) along with the assignments (\ref{CountAssignment}) imply that
\[
	\begin{split}
	F^2_{n,0}(0,r,\theta) &= r_{n,1}\sin(\theta_{n,1} - \theta_{n,0}) + r_{n,-1}\sin(\theta_{n,-1} - \theta_{n,0}) \\
	&\quad+ r_{n+1,0}\underbrace{\sin(\theta_{n+1,0} - \theta_{n,0})}_{= 0} + r_{n-1,0}\underbrace{\sin(\theta_{n-1,0} - \theta_{n,0})}_{= 0} \\
	&= \pm r_{n,1}\sin(\theta_{n,1}) \mp r_{n,-1}\sin(\theta_{n,-1})=0,
	\end{split}
\]
for all $n \geq 0$. This shows that our boundary conditions (\ref{Counter_Boundary}) necessarily give that $F^2_{n,0}(C,r,\theta) = 0$, and in turn will reduce the number of equations required to obtain a counterwinding vortex solution.

\begin{figure} 
	\centering
	\includegraphics[width=0.4\textwidth]{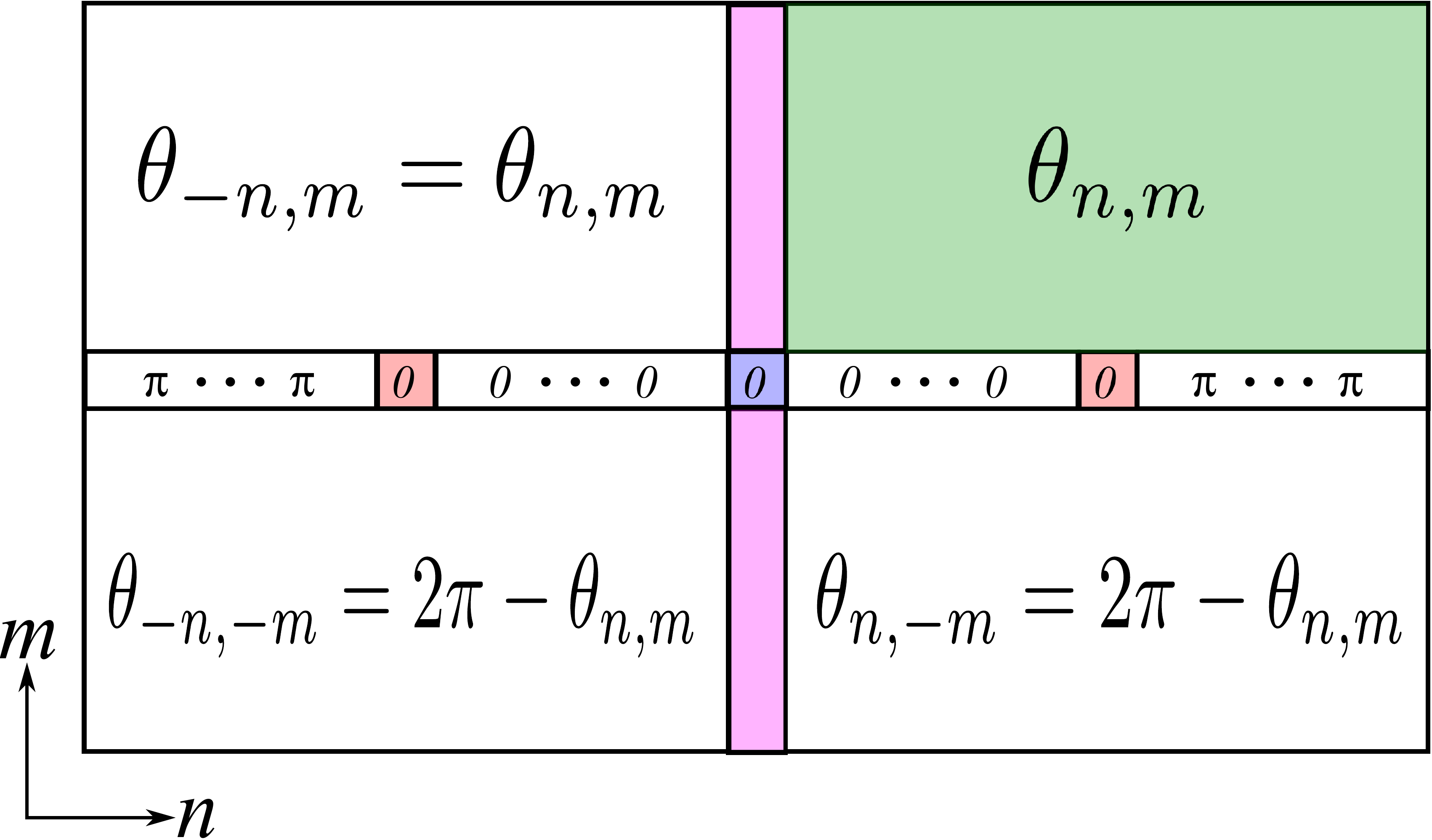} \quad \quad \quad \quad
	\includegraphics[width=0.4\textwidth]{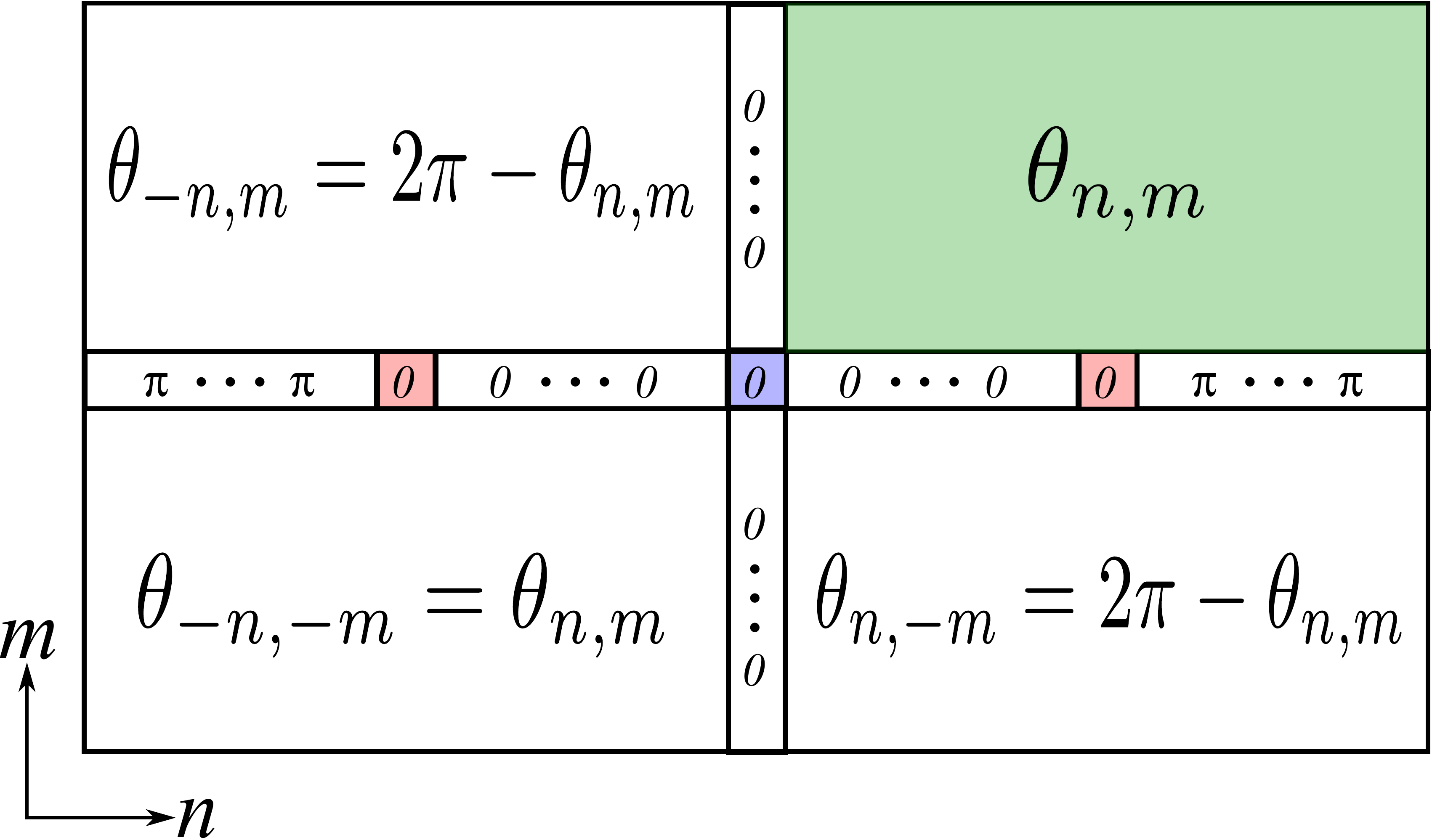}
	\caption{The symmetries of (left) counterwinding and (right) cowinding vortex solutions. The red shaded cells represent the indices $(n,m) = (\pm c,0)$ and the blue shaded cell is the center of the lattice $(n,m) = (0,0)$. The green shaded region represents the indices $n,m > 0$, and symmetry-based extensions beyond this region are indicated in each of the remaining three regions. The fuchsia region represents the cells with indices $(0,m)$, $m \neq 0$, which are not fixed by the symmetry of the counterwinding vortex solution.}
\label{fig:Vortex_Symmetry}
\end{figure}

Then, for a fixed $C>0$ solutions of $F^1(C,r,\theta) = F^2(C,r,\theta) = 0$ can be extended over the entire lattice through the following extension:
\[
	\begin{split}
		r_{-n,m} &= r_{n,m}, \quad
		r_{n,-m} = r_{n,m}, \quad
		r_{-n,-m} = r_{n,m}, \\
		\theta_{-n,m} &= \theta_{n,m}, \quad
		\theta_{n,-m} = 2\pi - \theta_{n,m}, \quad
		\theta_{-n,-m} = 2\pi - \theta_{n,m},
	\end{split}	
\]	
for each $n,m\geq 0$. The symmetries of the phase components over the full lattice are given on the left in Figure~\ref{fig:Vortex_Symmetry}, and we note that the symmetries of the radial components are significantly simpler since they are identical in each of the four regions of the figure. Notice that counterwinding vortex solutions necessarily have an $(n,m) \mapsto (-n,m)$ flip symmetry.

We may do something similar for cowinding vortices by introducing the boundary conditions
\begin{equation}\label{Co_Boundary}
	r_{-1,m} = r_{1,m}, \quad r_{n,-1} = r_{n,1}, \quad \theta_{-1,m} = 2\pi - \theta_{1,m}, \quad \quad \theta_{n,-1} = 2\pi - \theta_{n,1},
\end{equation}
for all $n,m\geq 0$. Similar to counterwinding case, these boundary conditions require that $\theta_{n,0} \in \{0,\pi\}$ for all $n \geq 0$, and hence for $c > 0$ as described above we have that
	\begin{equation}\label{CoAssignments}
		\theta_{0,m} = 0, \quad \theta_{n,0} = \left\{
     		\begin{array}{cl} 0, & n \leq c \\ \\
		\pi, & n > c.
		\end{array}
   	\right.
	\end{equation}
An important distinction between the counterwinding and cowinding vortices is that in the latter case the values of $\theta_{0,m}$ are fixed by the boundary conditions (\ref{Co_Boundary}), whereas in the counterwinding case we do not necessarily have explicit values for these phase components. Furthermore, the boundary conditions (\ref{Co_Boundary}) along with the assignments (\ref{CoAssignments}) imply that
\[
	\begin{split}
	F^2_{0,m}(0,r,\theta) &= r_{1,m}\sin(\theta_{1,m} - \theta_{0,m}) + r_{-1,m}\sin(\theta_{-1,m} - \theta_{0,m}) \\
	&\quad+ r_{0,m+1}\underbrace{\sin(\theta_{0,m+1} - \theta_{0,m})}_{= 0} + r_{0,m-1}\underbrace{\sin(\theta_{0,m-1} - \theta_{0,m})}_{= 0} \\
	&= r_{1,m}\sin(\theta_{1,m}) + r_{-1,m}\sin(\theta_{-1,m})=0,
	\end{split}
\]
for all $m \geq 0$, again due to the selection of boundary conditions. Similarly, the conditions for $m = 0$ give that $F^2_{n,0}(C,r,\theta) = 0$ for all $n \geq 0$, and therefore we are only left to solve $F^2_{n,m}(C,r,\theta) = 0$ for all $n,m > 0$ for $\{\theta_{n,m}\}_{n,m > 0}$. This shows that our assignments (\ref{CoAssignments}) necessarily give that $F^2_{0,m}(0,r,\theta) = F^2_{n,0}(C,r,\theta) = 0$ for all $n,m \geq 0$, and therefore reduces the number of equations required to obtain a cowinding vortex solution.

Then, for a fixed $C > 0$, solutions to $F^1(C,r,\theta) = F^2(C,r,\theta) = 0$ can be extended over the entire lattice through the following definitions:
\[
	\begin{split}
		r_{-n,m} &= r_{n,m}, \quad
		r_{n,-m} = r_{n,m}, \quad
		r_{-n,-m} = r_{n,m}, \\
		\theta_{-n,m} &= 2\pi - \theta_{n,m}, \quad
		\theta_{n,-m} = 2\pi - \theta_{n,m}, \quad
		\theta_{-n,-m} = \theta_{n,m},
	\end{split}	
\]	
for each $n,m\geq 0$. The symmetries of the phase components over the full lattice are given on the right in Figure~\ref{fig:Vortex_Symmetry}, and again we note that the symmetries of the radial components are significantly simpler since they are identical in each of the four regions of the figure. Notice that cowinding vortex solutions necessarily have an $(n,m) \mapsto (-n,-m)$ flip symmetry.

For both types of vortices considered in this work, the definition of the functions $F^1,F^2$ show that we may exploit the symmetries of the system (\ref{Polar_dNLS}) to greatly reduce the number of equations required to obtain a vortex solution. Most importantly, in the anti-continuum limit $C = 0$ we have
\[
	F^1_{n,m}(0,r,\theta) = r_{n,m}(1 - r_{n,m}^2),
\]
for all $n,m\geq 0$. Requiring that $r_{n,m}$ be nonnegative implies that $r_{n,m} \in \{0,1\}$ for all $n,m\geq 0$. In our case we will consider $r_{n,m} = 1$, for all $n,m\geq 0$ and $n \neq c$, along with the following two scenarios: $r_{c,0} = 0$ and $r_{c,0} = 1$.
The former case corresponds to the vortical configurations that
we will numerically consider below. The latter will be associated
with a complementary branch that will arise in the relevant bifurcation
diagram (cf. the details in the next subsection).
This will give two vortex solutions of each type to continue in $C \geq 0$, and also describes a process by which counterwinding vortices can be obtained numerically by restricting ourselves to a finite positive range of integers $n,m$, i.e., the first quadrant. Then, once we have obtained a numerical solution in the anti-continuum limit, we may continue this solution in $C$ to move beyond this limit and into a region of parameter space where obtaining solutions becomes significantly more complicated since we must solve for roots of both $F^1_{n,m}$ and $F^2_{n,m}$ for all $n,m\geq 0$. Most importantly, the above discussion shows that the curves of counter- and cowinding vortices continued in $C$ up from the anti-continuum limit $C=0$ will always retain the symmetries of Figure~\ref{fig:Vortex_Symmetry}.

\subsection{Existence of Stationary Solutions}\label{subsec:Existence} 

\begin{figure}
\begin{tabular}{cc}
\includegraphics[width=0.4\textwidth]{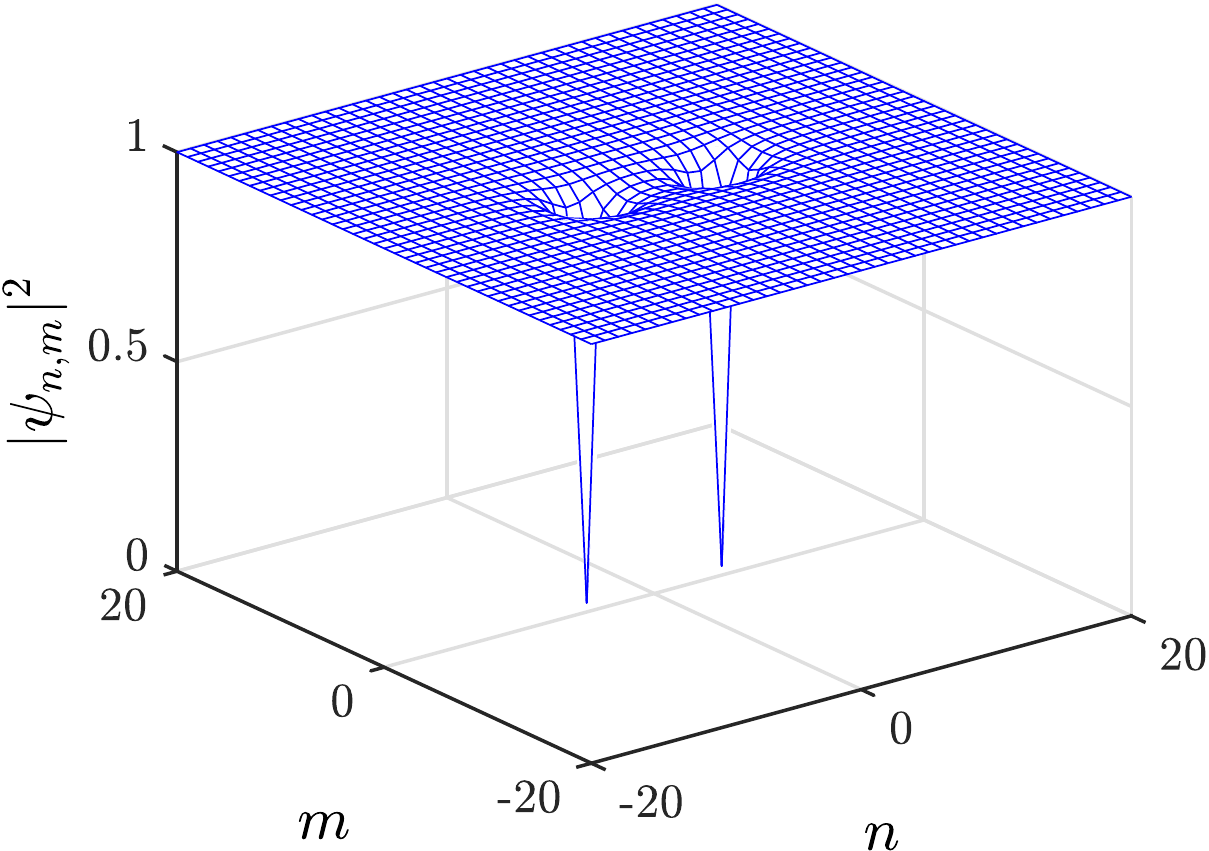} &
\includegraphics[width=0.4\textwidth]{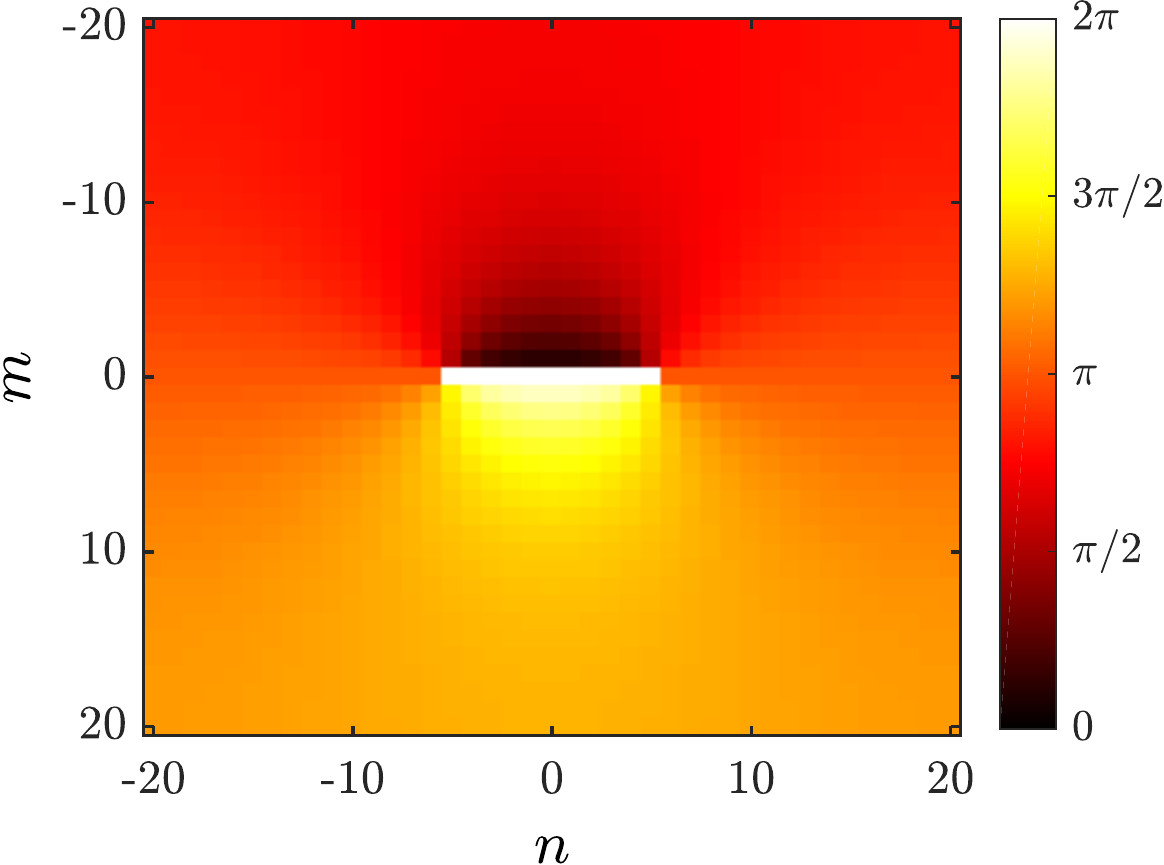} \\
\includegraphics[width=0.4\textwidth]{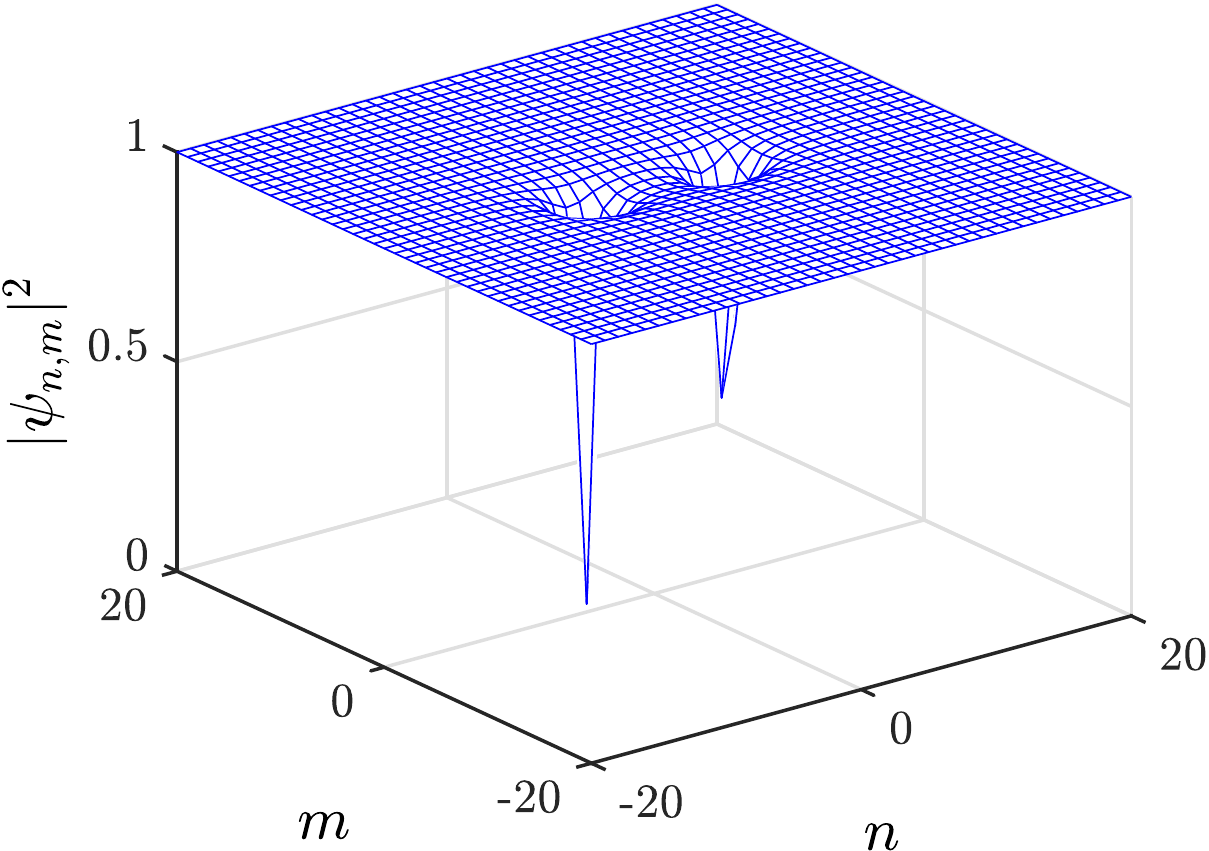} &
\includegraphics[width=0.4\textwidth]{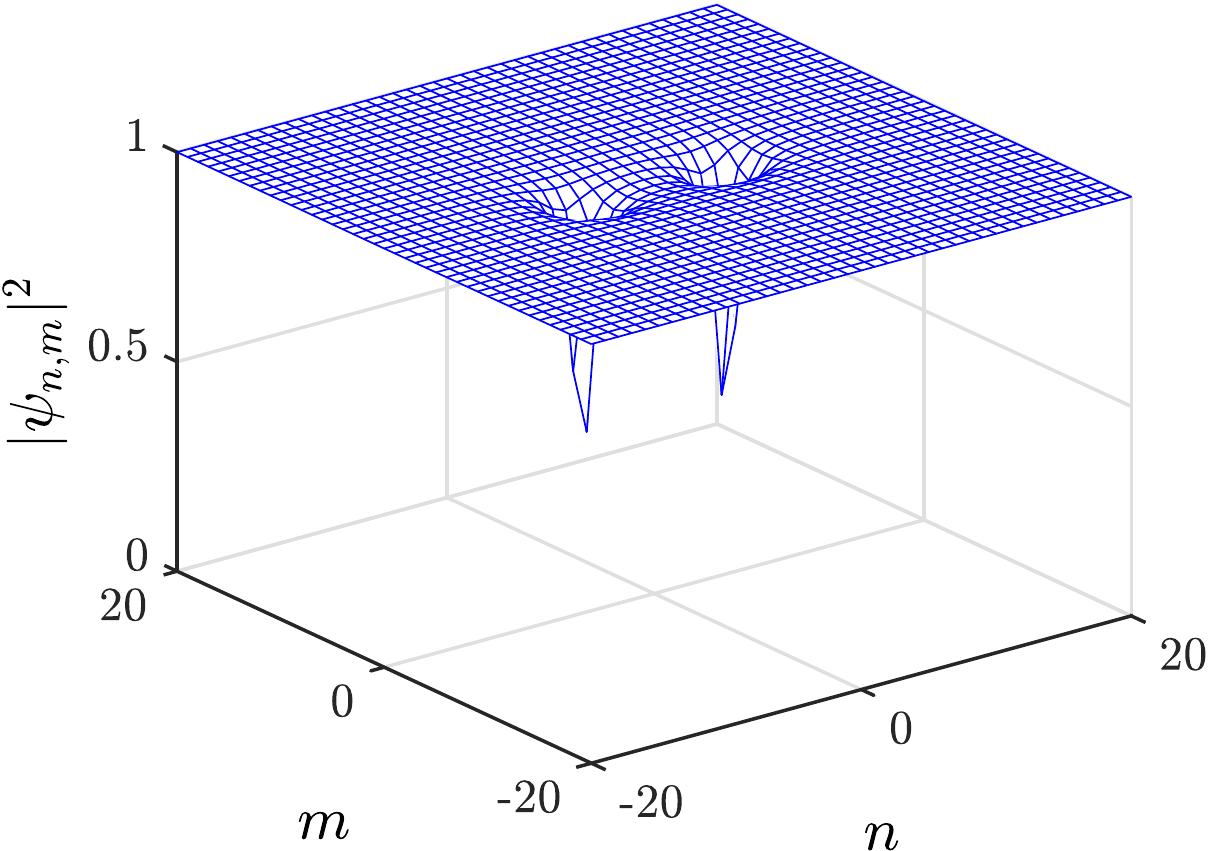}
\end{tabular}
\caption{{Counterwinding vortices for $C=0.4$ and $c=5$ in a $41\times41$ lattice. Density (top left panel) and phase (top right panel) of the symmetric 2VS (two-vortex states). Bottom panels show the density of 1VS (left) and 0VS (right). The phase of the latter solutions is not shown as they are almost identical to that of the 2VS.}}
\label{fig:profiles1}
\end{figure}

\begin{figure}
\begin{tabular}{cc}
\includegraphics[width=0.4\textwidth]{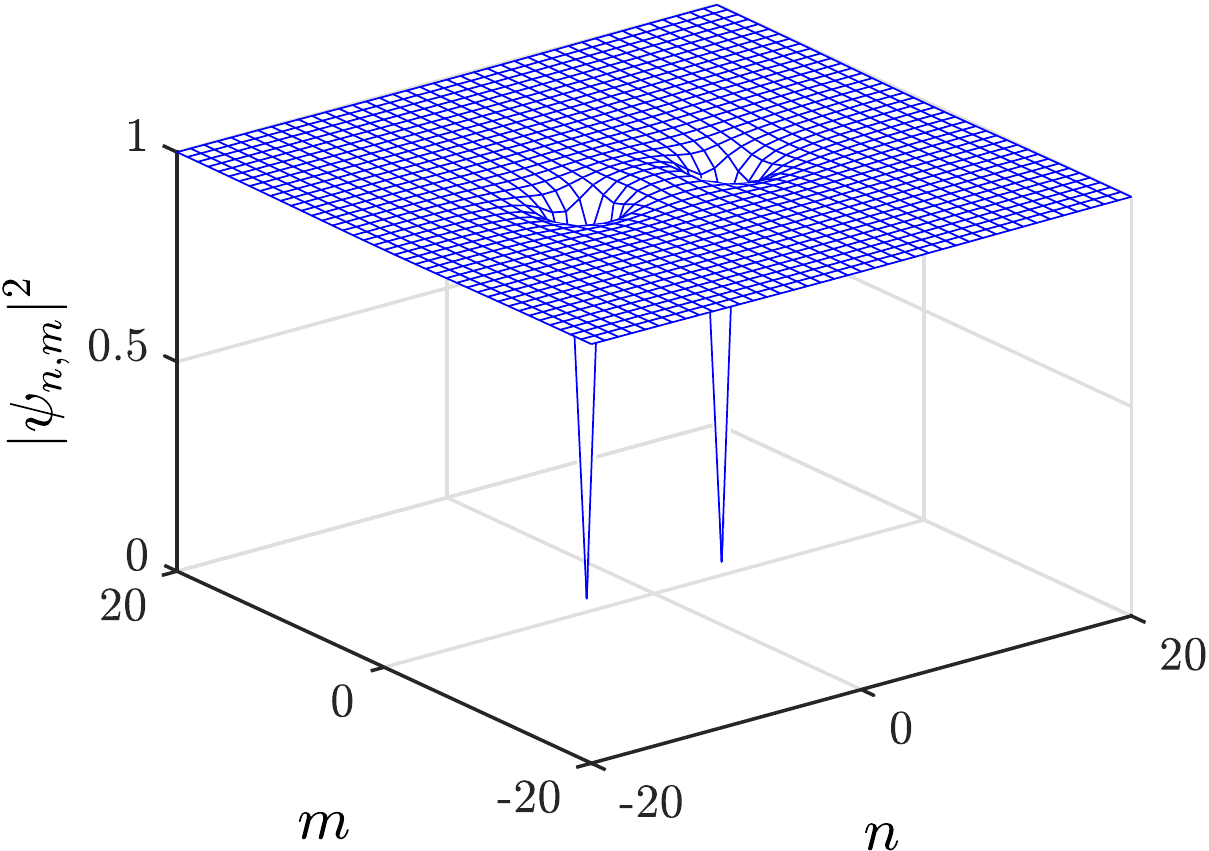} &
\includegraphics[width=0.4\textwidth]{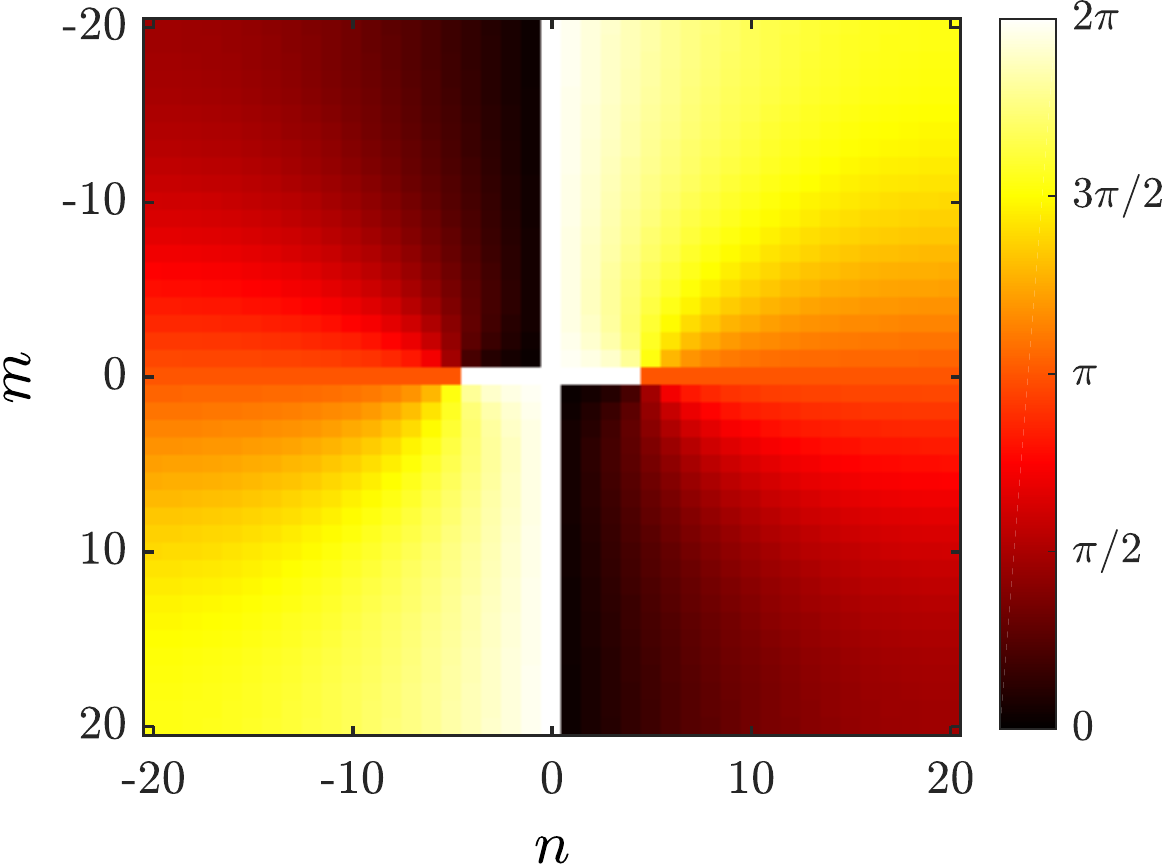} \\
\includegraphics[width=0.4\textwidth]{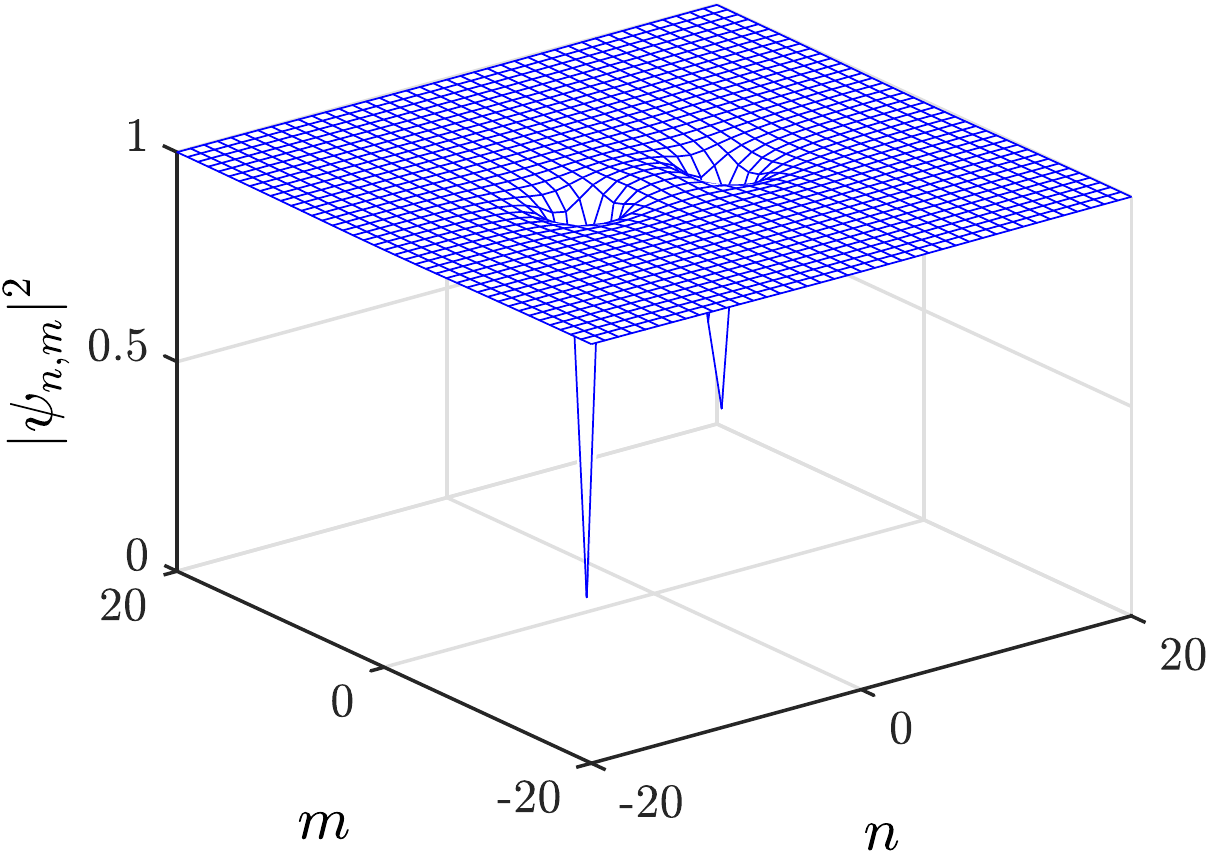} &
\includegraphics[width=0.4\textwidth]{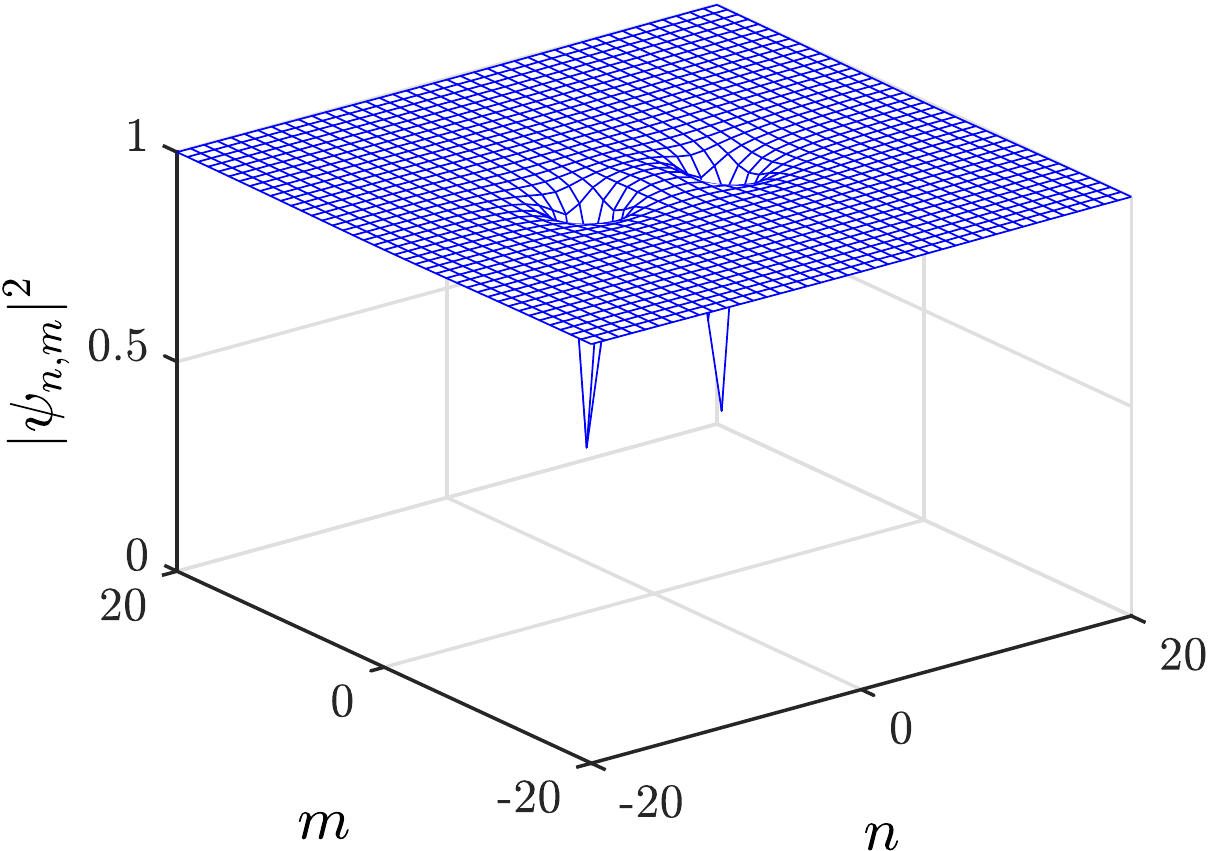} \\
\end{tabular}
\caption{{Same as Figure \ref{fig:profiles1} but for cowinding vortices}}
\label{fig:profiles2}
\end{figure}

Having set up the relevant existence problem of a stationary vortex
pairs analytically, we now turn to the corresponding numerical
considerations in $N\times N$ lattices. We will mostly present
numerical results for $N = 41$, but remark that our results have also
been checked with $N = 81$ and $251$ to determine
consistency. Moreover, unless otherwise stated, on every lattice we
take $c = 5$ but comment on the effect of changing $c$ towards the end
of this section. Using an ansatz such as the one described in the
previous section we are able to identify solutions involving two
distinct vortices of each type in the anti-continuum limit given by
fixing $r_{n,m} = 1$, for all $n,m\in \{0,\dots,(N-1)/2\}$ and $n \neq
5$. The vortex pair branch involves the selection of $r_{\pm 5, 0}=0$
in the anti-continuum limit of $C=0$; we are able to continue these relevant wave forms into $C > 0$. Examples of these continued solutions, hereby referred to as stationary symmetric counterwinding (cowinding) vortices, are depicted in Figure~\ref{fig:profiles1} (Figure~\ref{fig:profiles2}) on a $41\times41$ lattice for the parameter value $C = 0.4$.
In the figure structures involving zero or one vortex at
the anti-continuum limit are also shown; these structures
are explained in detail below.

\begin{figure}
\begin{tabular}{cc}
\includegraphics[width=0.4\textwidth]{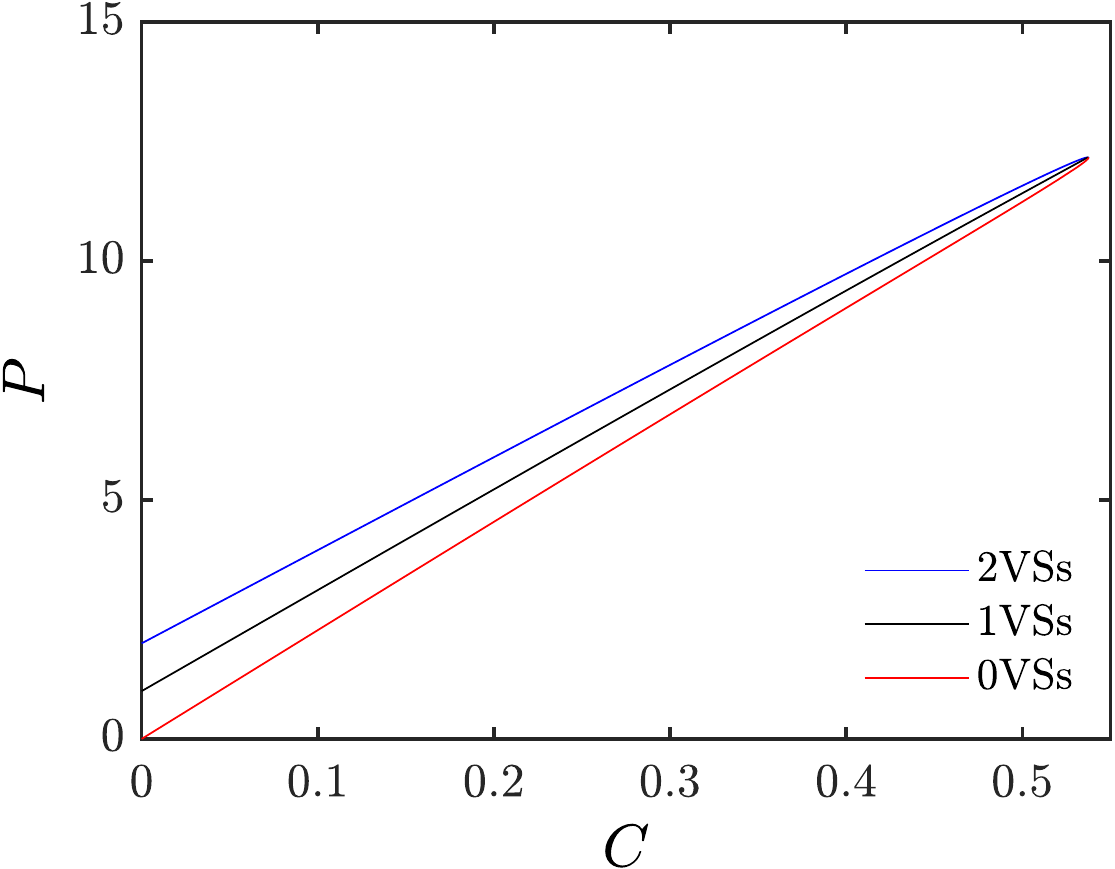} &
\includegraphics[width=0.4\textwidth]{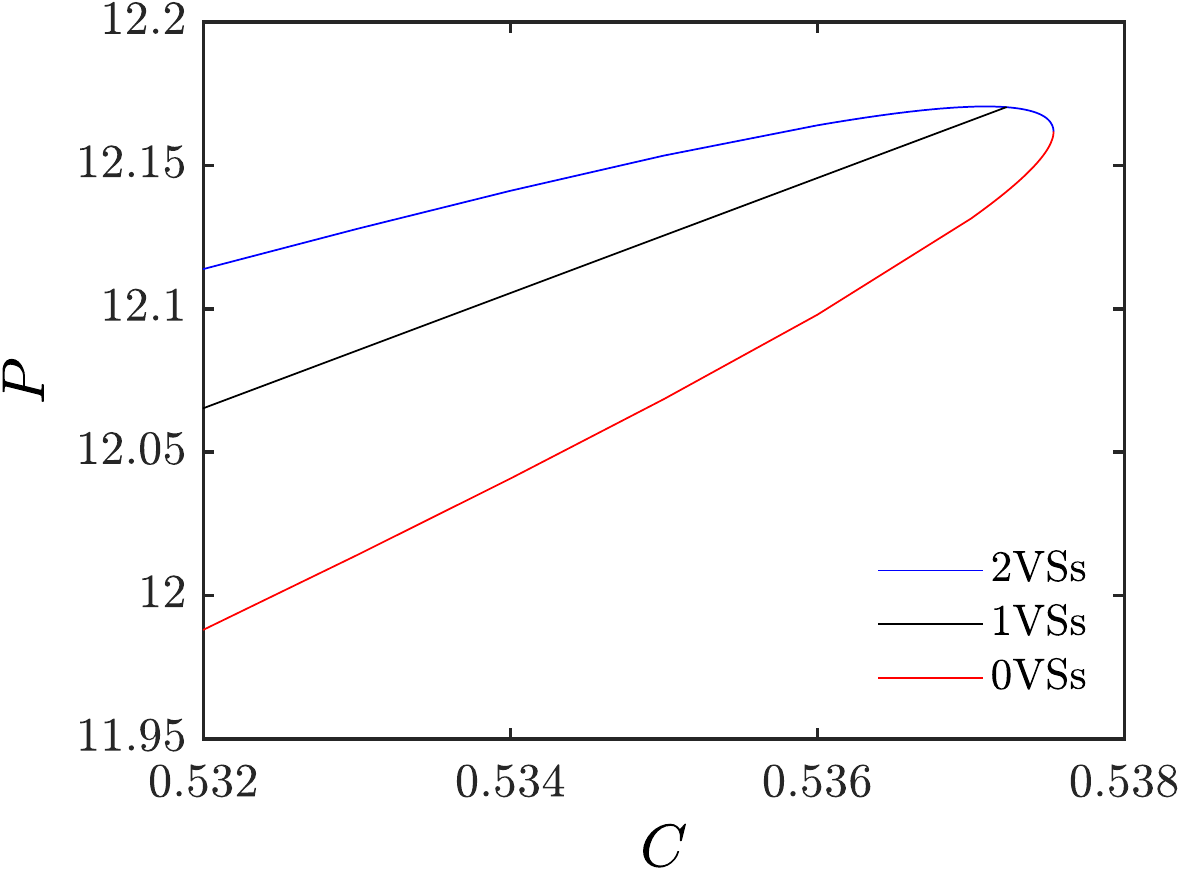} \\
\includegraphics[width=0.4\textwidth]{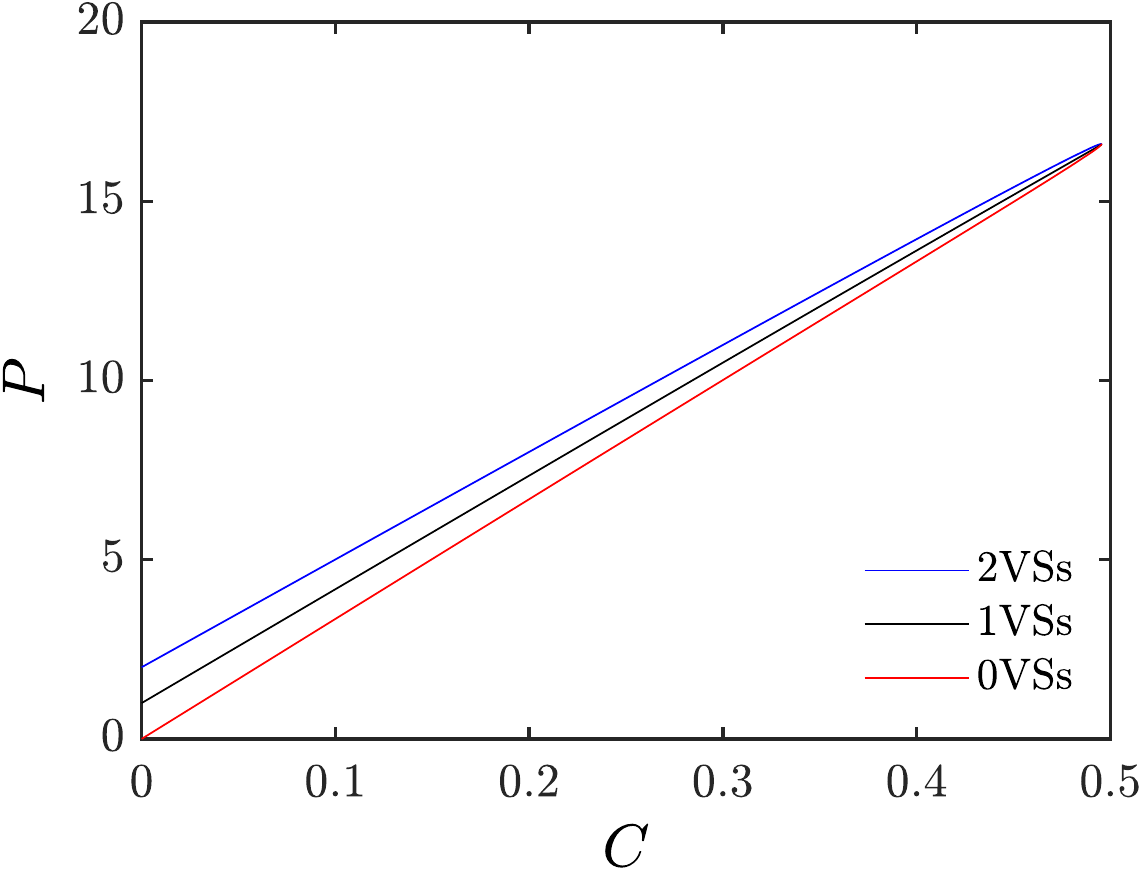} &
\includegraphics[width=0.4\textwidth]{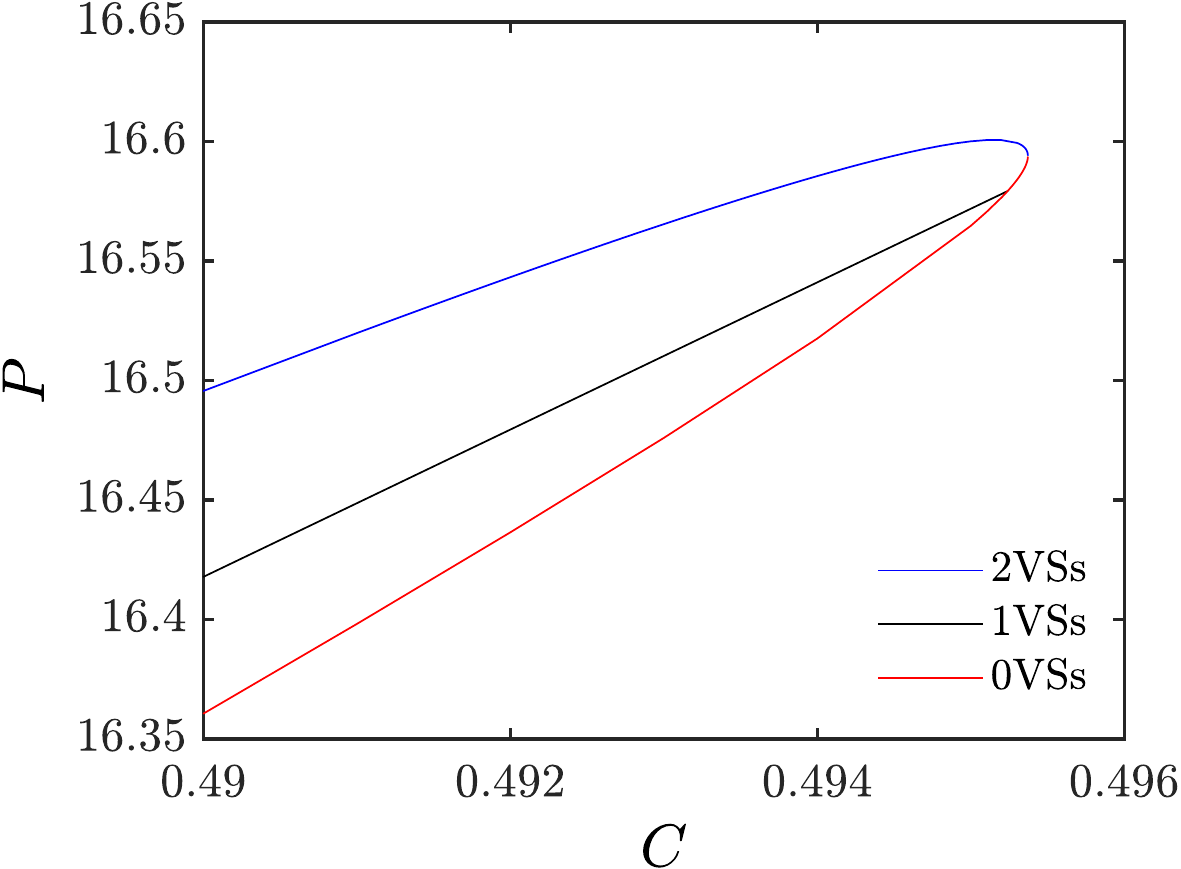} \\
\end{tabular}
\caption{Dependence of the complementary norm $P$ versus the coupling constant $C$ for (top) counterwinding and (bottom) cowinding vortices with $c=5$ on a $41\times41$ lattice. The right panels are a zoom in of the turning point and pitchfork bifurcations.}
\label{fig:Continuation}
\end{figure}

\begin{figure}
\begin{tabular}{cc}
\includegraphics[width=0.4\textwidth]{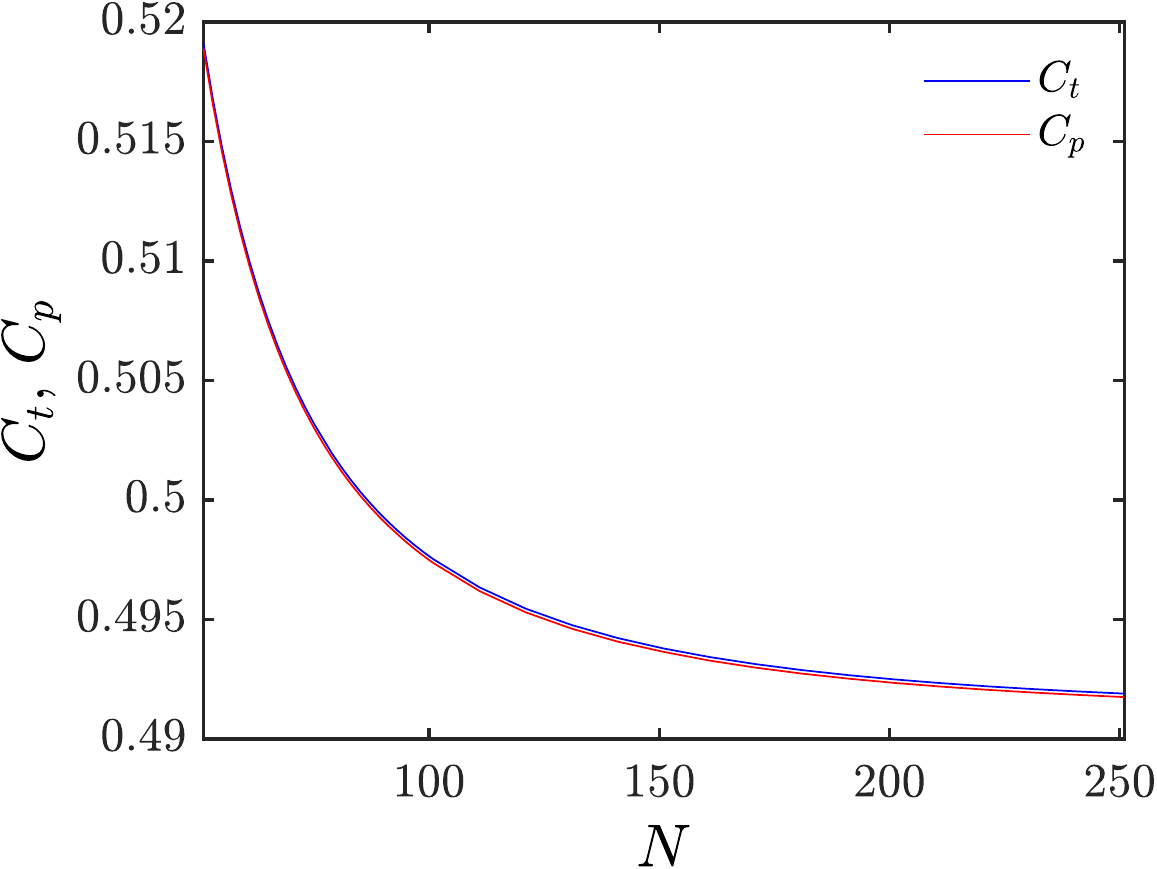} &
\includegraphics[width=0.4\textwidth]{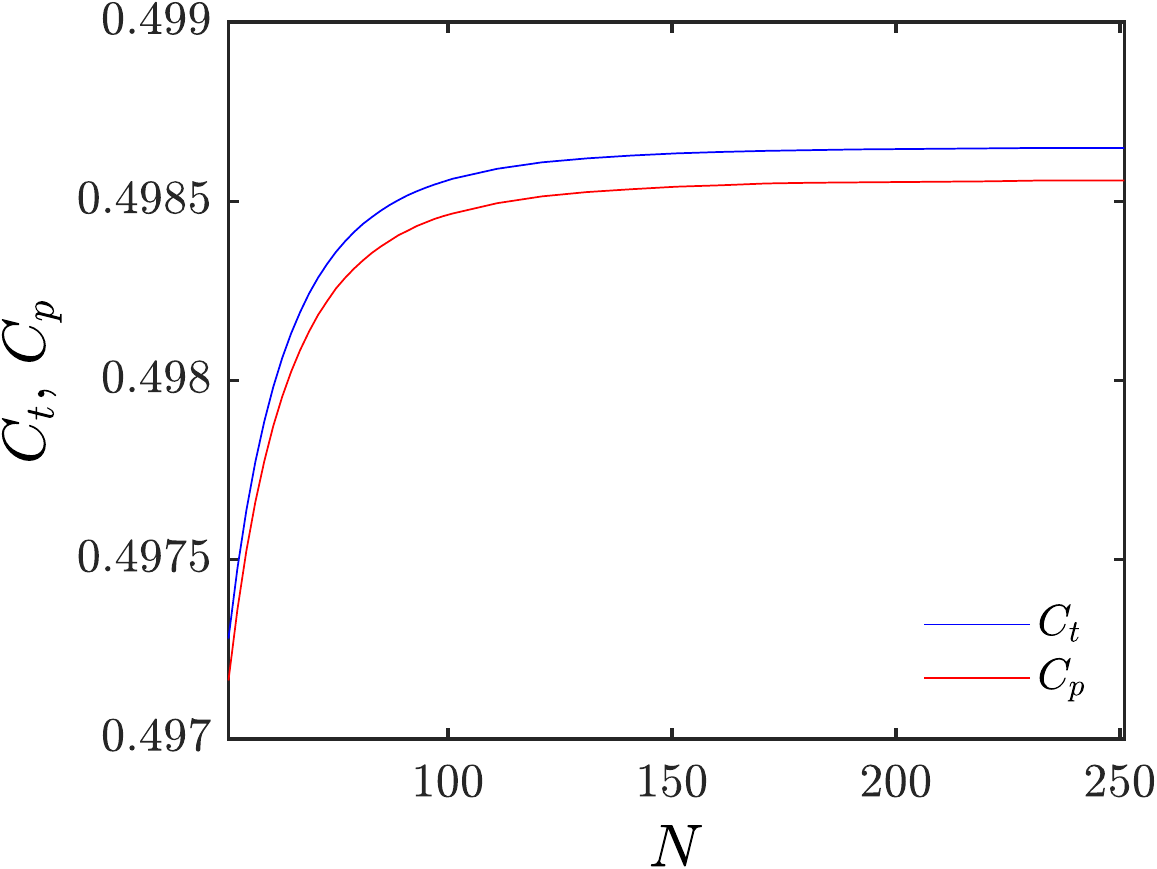} \\
\end{tabular}
\caption{{Dependence of the bifurcation points $C_p$ and $C_t$ with respect to the lattice size $N$ for counterwinding (left panel) and cowinding vortices (right panel).}}
\label{fig:bifsN}
\end{figure}

Contrary to single vortices that can be continued throughout the interval of real non-negative values of $C$, we find that stationary
symmetric counter- and cowinding vortices do not exist for all values of the coupling constant $C$.
Of course, this is natural to expect given the absence of such stationarity in the continuum limit. However, the interest in our case involves the transition from the anti-continuum stationarity to the continuum traveling.
We depict the bifurcation diagram of the stationary counter- and cowinding vortices on a lattice with $N = 41$ in Figure~\ref{fig:Continuation} with the vertical axis given by the complementary norm
\begin{equation}\label{PNorm}
	P = \sum_n\sum_m (1 - |\phi_{n,m}|^2),
\end{equation}
where we recall that $1$ is the background density. For both types of vortices we have $P$ is equal to $0$ and $2$ in the anti-continuum limit, depending on
whether $r_{\pm c,0}=1$ or whether {$r_{\pm c,0} = 0$}.
We will refer to the continued solutions in $C>0$ as a 0VS and a 2VS, respectively, i.e., as bearing $0$ or $2$ vortices, respectively. It is important to highlight that
the relevant terminology is principally meaningful in the anti-continuum limit, yet
by extension in the manuscript, we will refer
to the branches using the same notation for non-vanishing values of $C$.
The upper branches in blue in Figure~\ref{fig:Continuation} correspond
to a stationary 2VS with $r_{\pm5,0} = 0$ in the anti-continuum limit,
whereas the lower branches in
red correspond to a stationary 0VS with $r_{\pm5,0} = 1$ in the anti-continuum limit. Our numerics reveal that upon continuing these solutions into $C > 0$, $r_{\pm5,0}$ monotonically increases with $C$, whereas 0VSs have $r_{\pm5,0}$ monotonically decreasing as $C$ increases. This monotonic decrease eventually terminates at a turning point bifurcation $C = C_t$ where these two symmetric vortex states (the 2VS and the 0VS) collide and annihilate each other. Our numerical investigations have revealed that this scenario is independent of the number of lattice sites, but we do remark that the exact value at which the turning point takes place does appear to change with $N$. In particular, we have found that on the $41\times41$ lattice we have $C_t = 0.5375386$ for counterwinding vortices and $C_t = 0.4953718$ for cowinding vortices.
For counterwinding (cowinding) vortices the relevant critical point location slowly decreases (increases) as a function of increasing lattice size $N$ up to an asymptotic value, as shown in Figure~\ref{fig:bifsN}.

\begin{figure}
\includegraphics[width=0.4\textwidth]{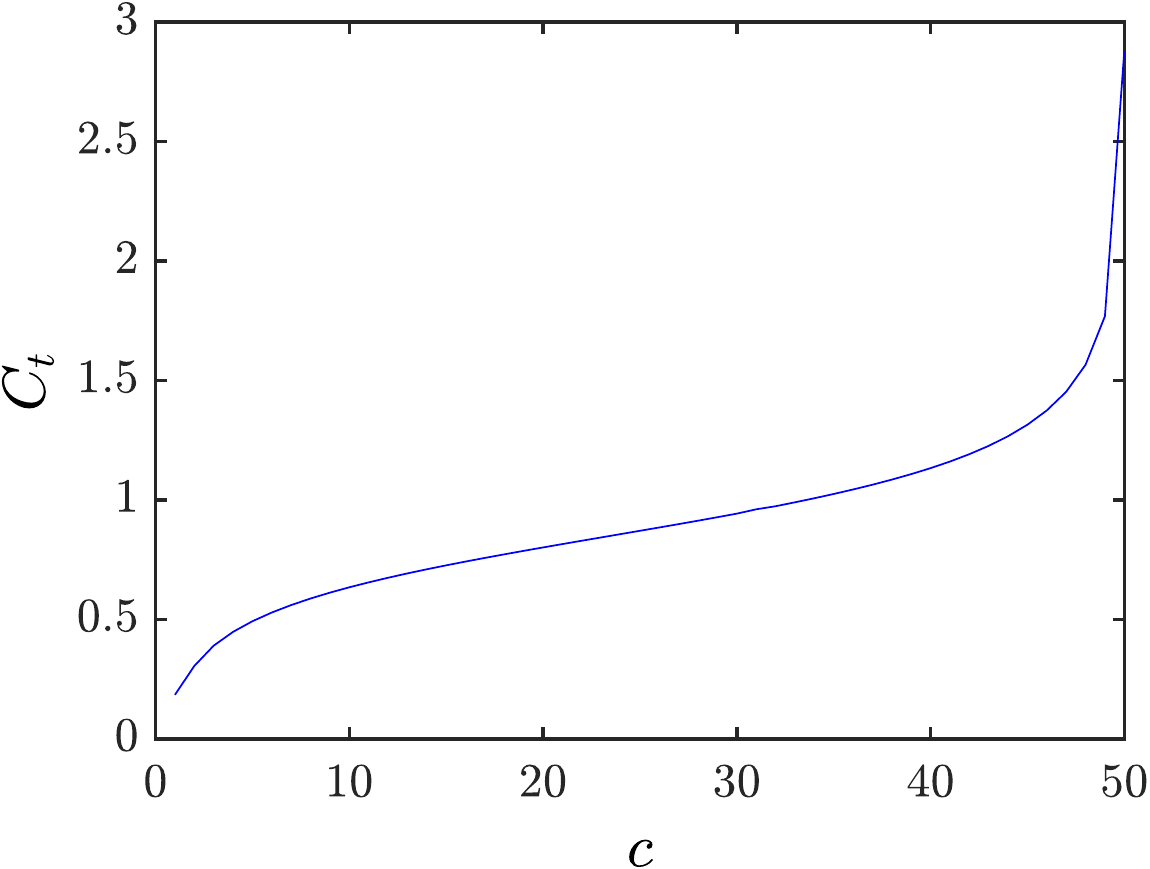}
\caption{Dependence of the bifurcation points $C_t$ with respect to $c$ for counterwinding vortices with $N=201$. Notice the sharp increasing when $c=50$.}
\label{fig:bifsd}
\end{figure}

Our investigation has revealed that there exists another pair of each vortex type which cannot be obtained via the functions $F^{1,2}$ since they do not satisfy the symmetries of Figure~\ref{fig:Vortex_Symmetry}. In the anti-continuum limit these solutions are characterized by taking $r_{n,m} = 1$ for all $(n,m) \neq (\pm5,0)$, with $r_{5,0} = 1 - r_{-5,0} \in \{0,1\}$, and therefore we will hereby refer to these asymmetric cowinding vortices as 1VSs since their value in the complementary norm (\ref{PNorm}) is exactly 1 in the anti-continuum limit (and they effectively involve only one vortex instead of two). An example counterwinding vortex profile is given in the bottom left panel of Figure~\ref{fig:profiles1} and analogously we provide a sample cowinding 1VS in Figure~\ref{fig:profiles2}. We find that continuing these solutions up from the anti-continuum limit leads to one of $r_{5,0}, r_{-5,0}$ increasing monotonically up from $0$ and the other decreasing monotonically down from $1$.
As is demonstrated in Figure~\ref{fig:Continuation}, these asymmetric counterwinding (cowinding) vortices bifurcate through a subcritical (supercritical) pitchfork bifurcation from the 2VS (0VS) branch of symmetric counterwinding vortices and the 0VS branch of symmetric cowinding vortices. This phenomenology is present irrespectively of lattice size but the value of $C_p$ does in fact vary with $N$, as can be observed in Figure~\ref{fig:bifsN}). Furthermore, these asymmetric solutions exist for all $C \in [0,C_p]$ starting from the anti-continuum limit and are mapped into each other by taking
\[
\begin{split}
	&{\rm Counterwinding:}\quad \phi_{n,m} \mapsto \phi_{-n,m}, \\
	&{\rm Cowinding:}\quad \phi_{n,m} \mapsto \phi_{-n,-m},
\end{split}
\]
for any $C \geq 0$ for which they exist, as is natural for two branches emerging as a result of a pitchfork bifurcation. We find that, as expected, the dependence of $C_t$ and $C_p$ with respect to the distance between vortices for fixed $N$ gives a monotonic increment of $C_t$ when the distance is increased. Figure~\ref{fig:bifsd} shows this phenomenon for counterwinding vortices by depicting $C_t$ versus $c$; as $C_t-C_p\lesssim10^{-4}$, the curve $C_p(c)$ is almost indistinguishable from $C_t(c)$ and we have decided not to include it. A monotonic trend for the critical point as a function of $c$ is also obtained in the cowinding case (results not shown here).

\subsection{Linear Stability}\label{subsec:Stability} 

The spectral stability of stationary solutions is obtained by means of Bogoliubov-de Gennes spectral linearization analysis. More specifically, the relevant ansatz of the form
\[
	\psi_{n,m}(t) = \sqrt{\omega}[\psi_{n,m} + \delta(p_{n,m}\mathrm{e}^{\lambda t} + q^*_{n,m}\mathrm{e}^{\lambda^*t})]\mathrm{e}^{-\mathrm{i}\omega t},
\]
is introduced into the differential equation (\ref{dNLS}). Then, at lowest order in $\delta$ the linear problem can be written as the eigenvalue problem:
\[
	\lambda\begin{pmatrix}
		p_{n,m} \\ q_{n,m}
	\end{pmatrix} = \mathrm{i}\begin{pmatrix}
		2|\phi_{n,m}|^2 - 1 - \frac{C}{2}\Delta & \phi^2_{n,m} \\
		-(\phi^2_{n,m})^* & 1 - 2|\phi_{n,m}|^2 + \frac{C}{2}\Delta
	\end{pmatrix}\begin{pmatrix}
		p_{n,m} \\ q_{n,m}
	\end{pmatrix}.
\]
As in the single vortex case of \cite{1Vortex}, there will be
eigenvalues with negative Krein signature/energy (the latter being defined as
$K=\sum_{n,m} |p_{n,m}|^2 - |q_{n,m}|^2$) hereby denoted as NEEs, as
well as  continuous spectrum, which of course will be discretized since we are numerically identifying these vortices on a finite lattice. At the anti-continuum limit nVSs of all types, with $n = 0,1,2$, have exactly $n$ pairs of degenerate (between them) NEEs with $\lambda=\pm\mathrm{i}$ corresponding to excited sites, and $N^2-2n$ eigenvalues with $\lambda = 0$ corresponding to the non-excited sites.

We begin with a discussion of counterwinding vortices. Moving into $C
> 0$ we find that the degeneracy of $\lambda = 0$ eigenvalues (and of
NEEs for 2VSs) is broken so that the continuous bands on the imaginary
axis become bounded away from the origin of the complex plane. As
illustrated also in the case of the single DNLS vortex~\cite{1Vortex}
(which, at the same time, is based in the analysis for 1D dark
solitons performed in \cite{johkiv}), the background nodes lead, for
finite $C$, to a continuous spectrum extending over the interval
$\lambda \in \mathrm{i}[-\sqrt{16C^2 + 8C},\sqrt{16C^2 + 8C}]$ along
the imaginary axis. At the same time, the absolute value of the NEEs
decreases
{in a quasi-linear way}, and we find that there exists a critical
value, denoted as {$C_c\approx0.080$}, for which one of them enters
the continuous band, creating a cascade of Hamiltonian Hopf and
inverse Hamiltonian Hopf bifurcations, thus leading to oscillatory
instabilities. The dependence of the NEEs for $C<C_c$ is almost the
same, up to a $\sim10^{-4}$ difference, between 2VS and 1VS. Following
the analysis performed in \cite{1Vortex}, the NEEs can be approximated
by $\lambda\approx\pm\mathrm{i}(1-2C)$, which leads to a collision
with the continuum band at $C=(2\sqrt{3}-3)/6\approx0.077$,
in good agreement with the numerical result. As in the single vortex case, due to the inverse Hamiltonian Hopf bifurcation there will be linearized stability windows that would not appear if the continuous spectrum were dense. In the limit $N \to \infty$ we expect to find that the vortex would be oscillatorily unstable whenever $C > C_c$. Notice that the value of $C_c$ decreases when the distance between vortices is decreased; in fact, for $c=1$, $C_c\approx0.071$.

\begin{figure}
\begin{tabular}{ccc}
\includegraphics[width=6cm]{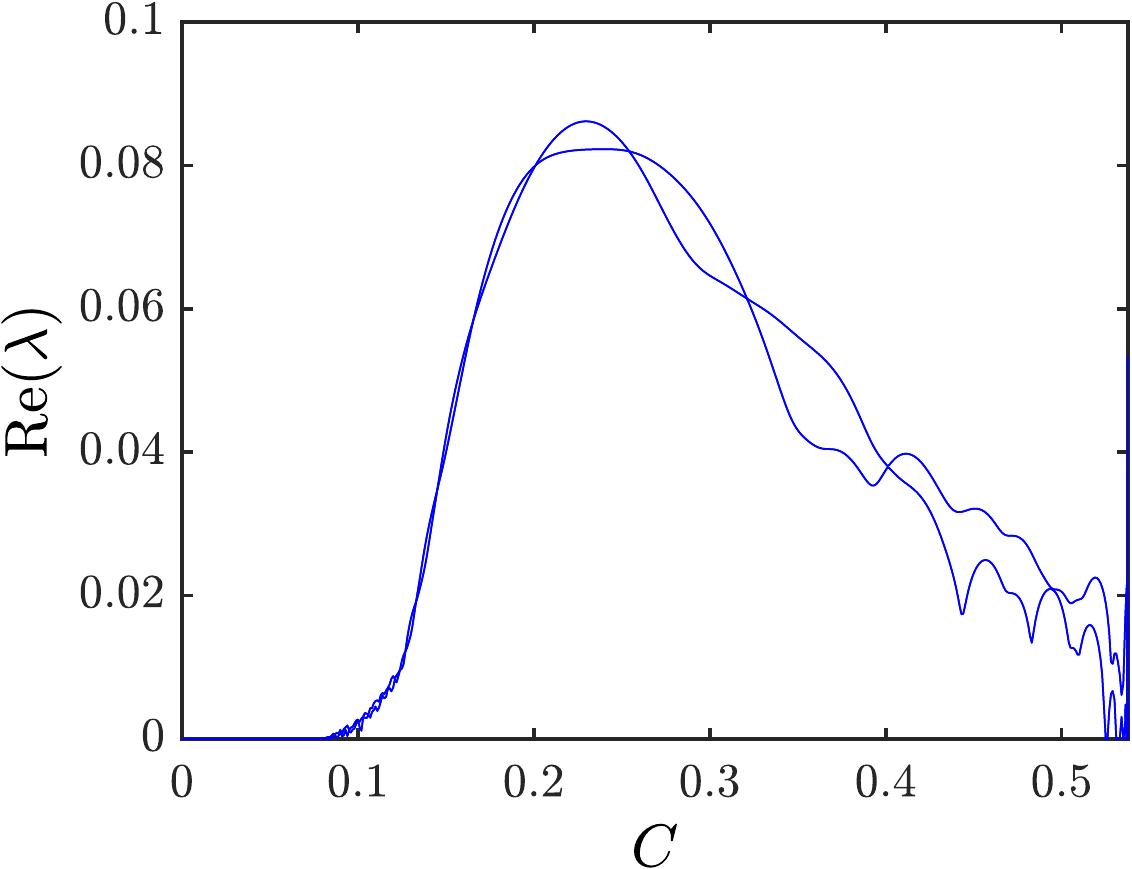} &
\includegraphics[width=6cm]{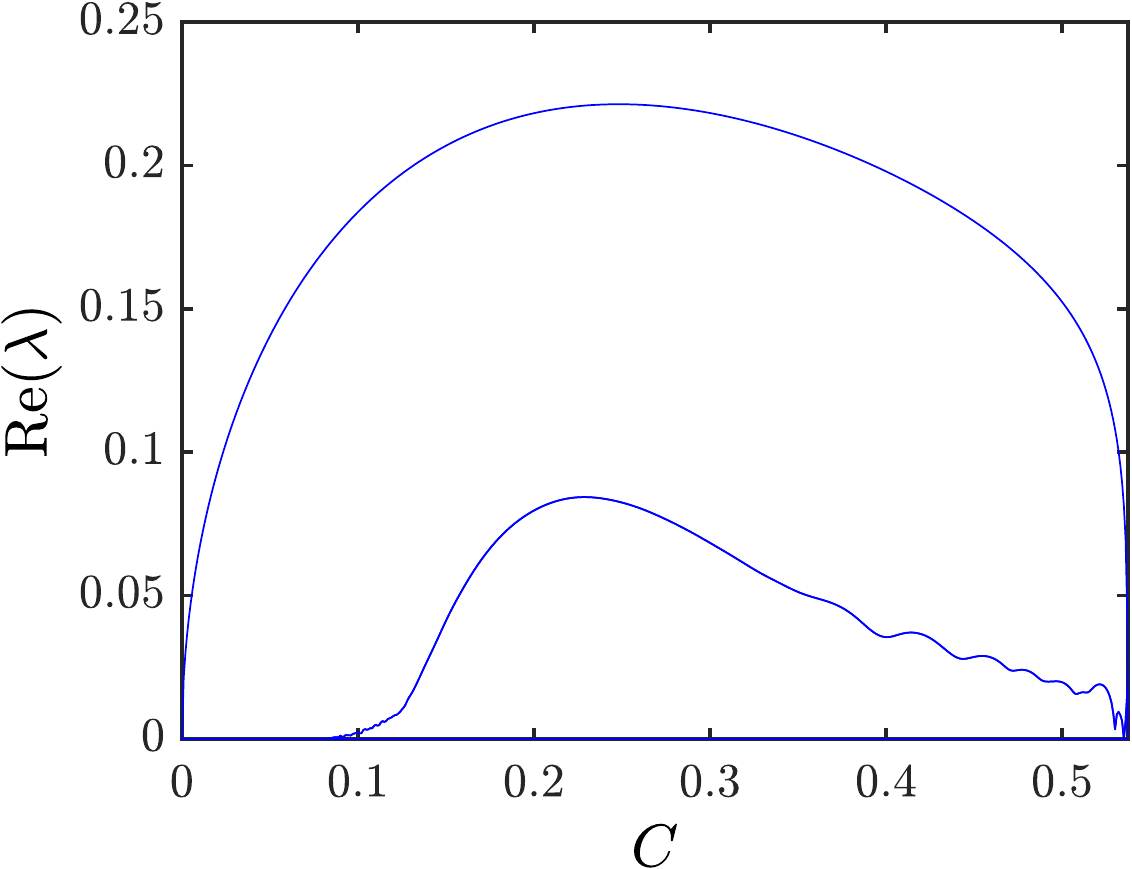} &
\includegraphics[width=6cm]{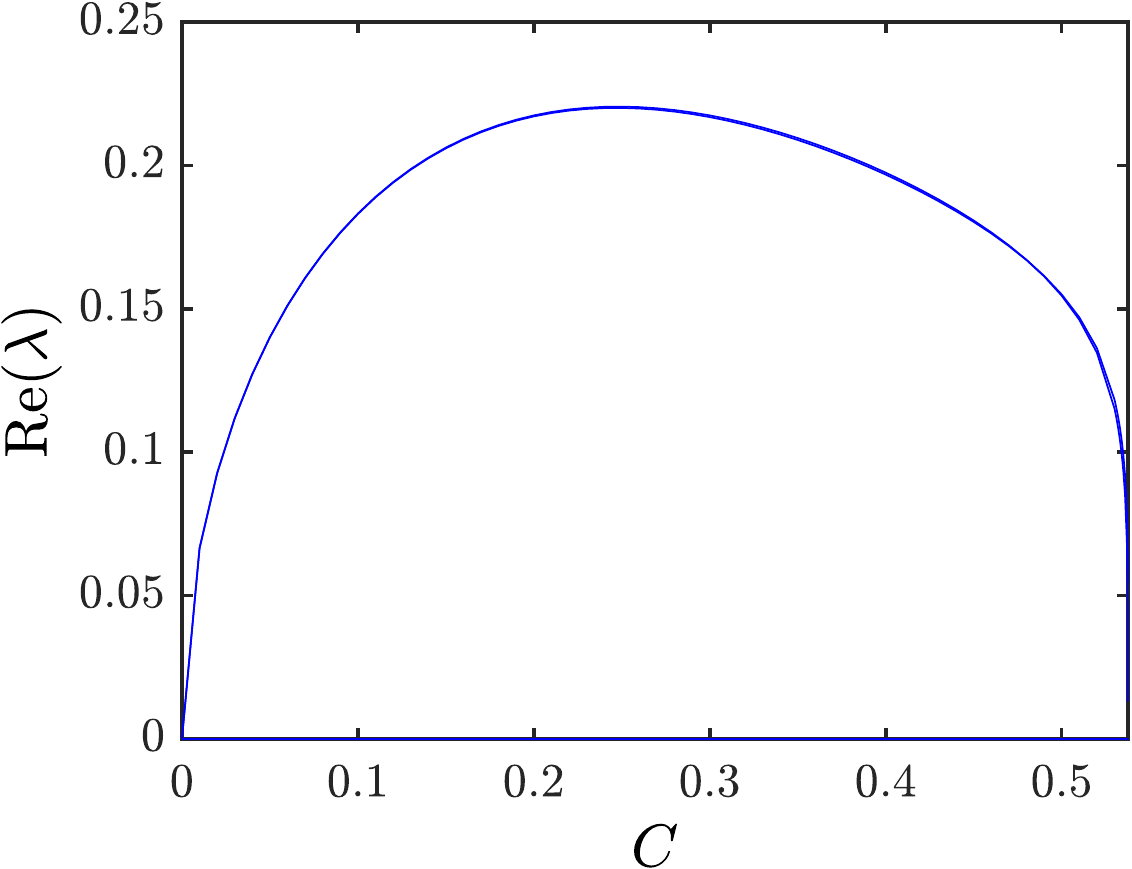} \\
\includegraphics[width=6cm]{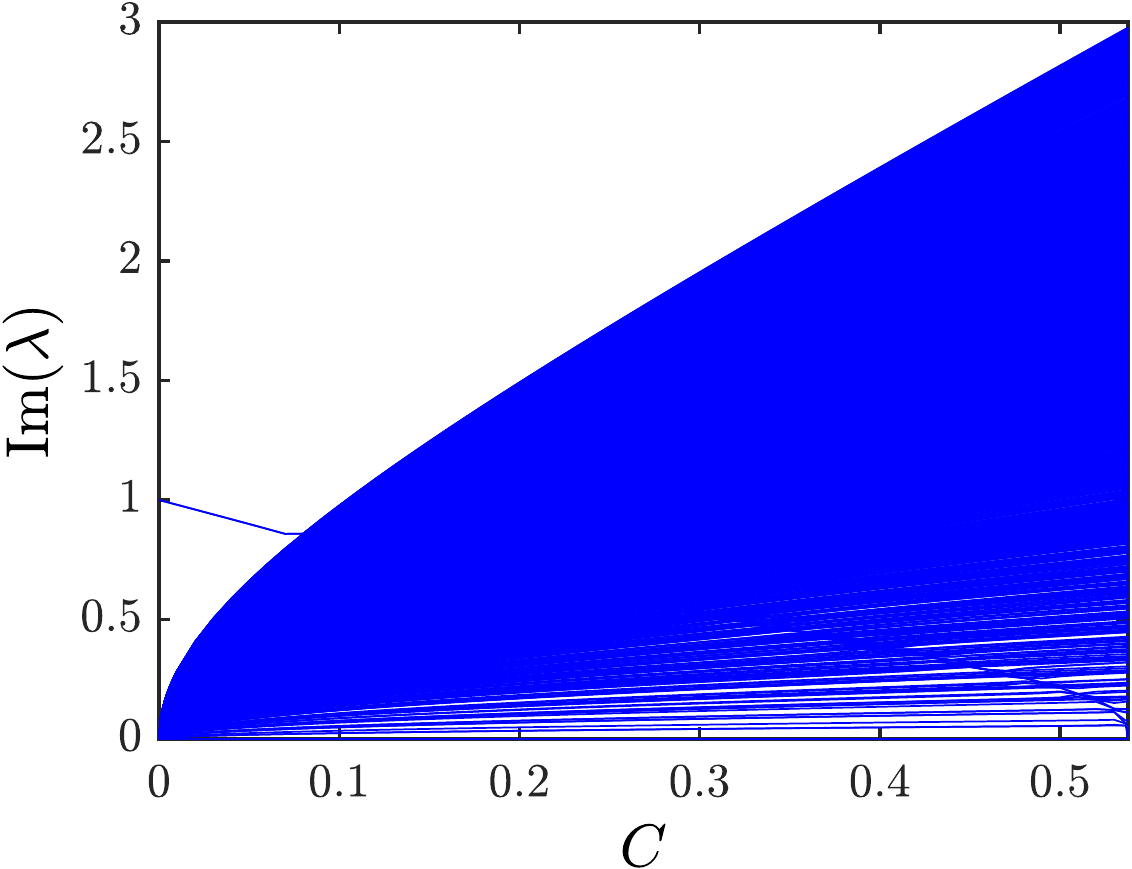} &
\includegraphics[width=6cm]{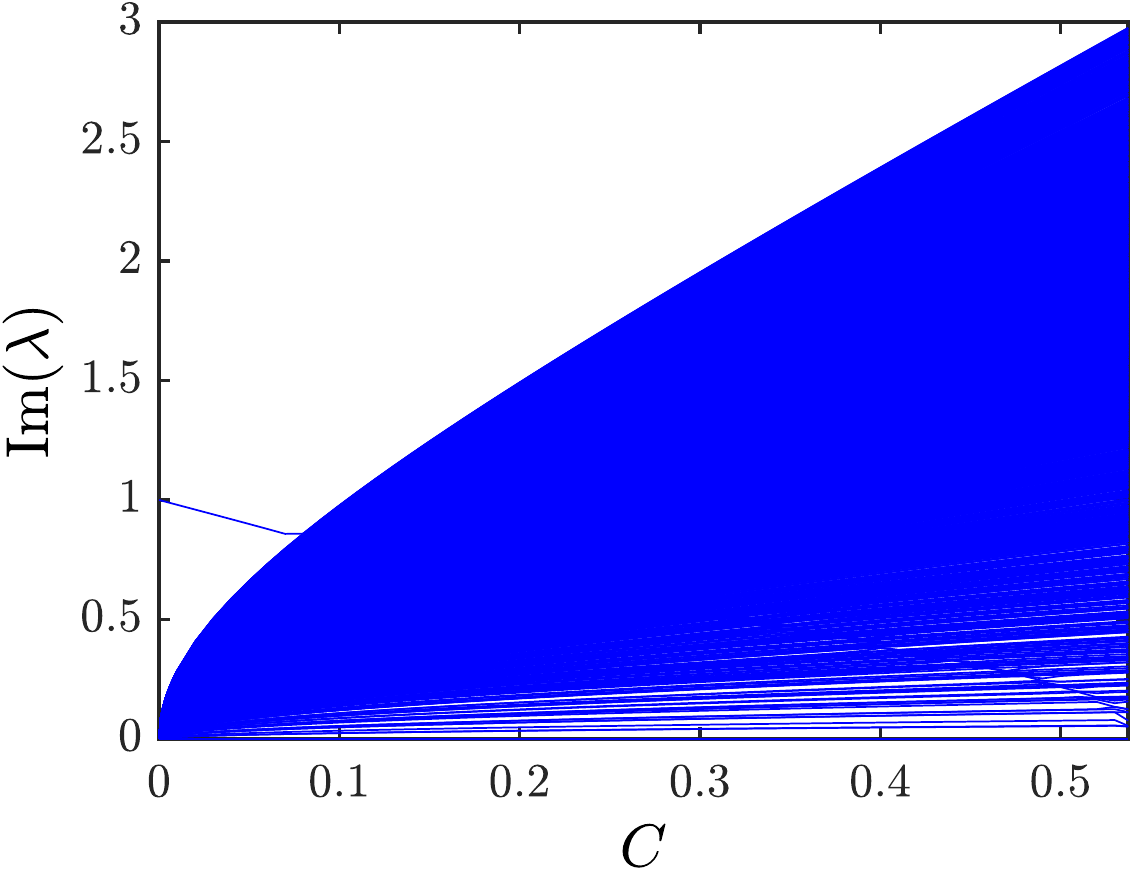} &
\includegraphics[width=6cm]{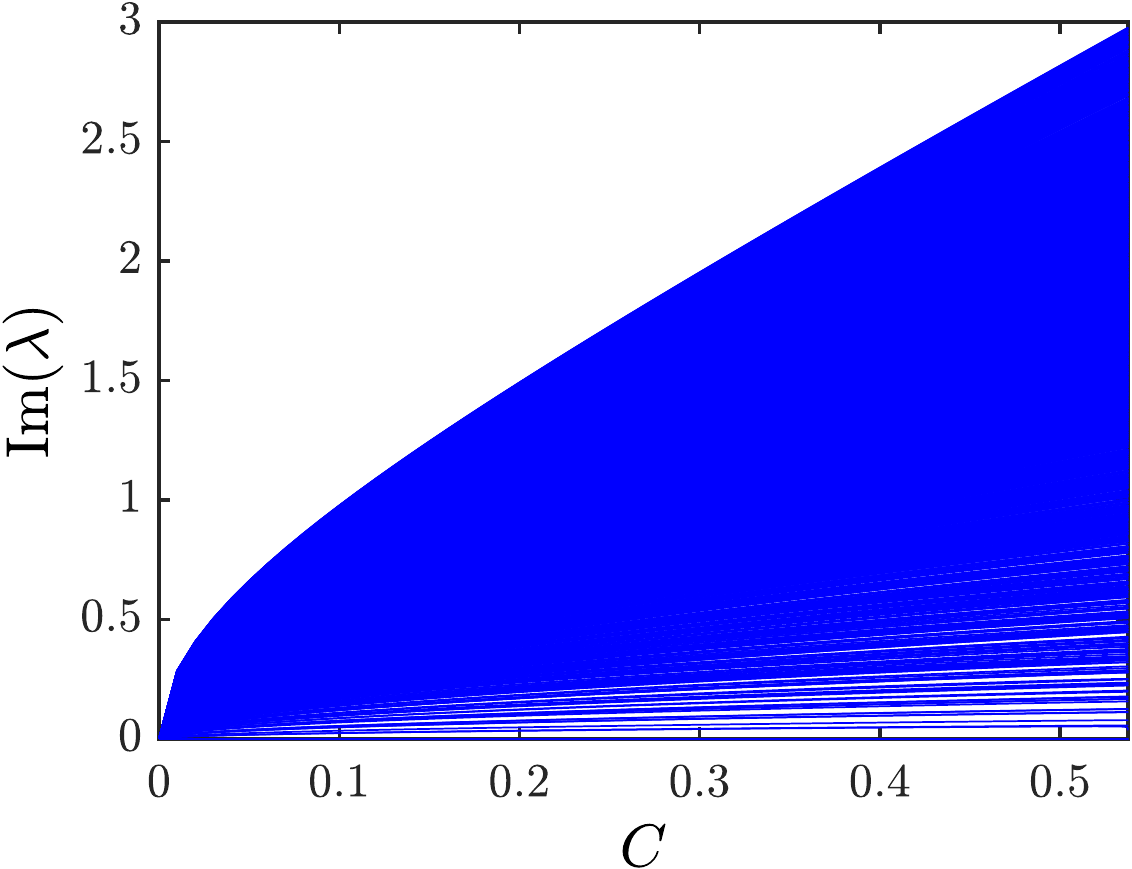}
\end{tabular}
\caption{Dependence with respect to $C$ of the (top) real and (bottom) imaginary parts of the eigenvalues of counterwinding (left) 2VSs, (middle) 1VSs, and (right) 0VSs. In every case, a $41\times41$ lattice has been used. Notice that among the 3 branches only the 2VS branch is stable up to a critical $C_c$ away from the anti-continuum limit of $C=0$.}
\label{fig:stab1}
\end{figure}

\begin{figure}
\begin{tabular}{cc}
\includegraphics[width=6cm]{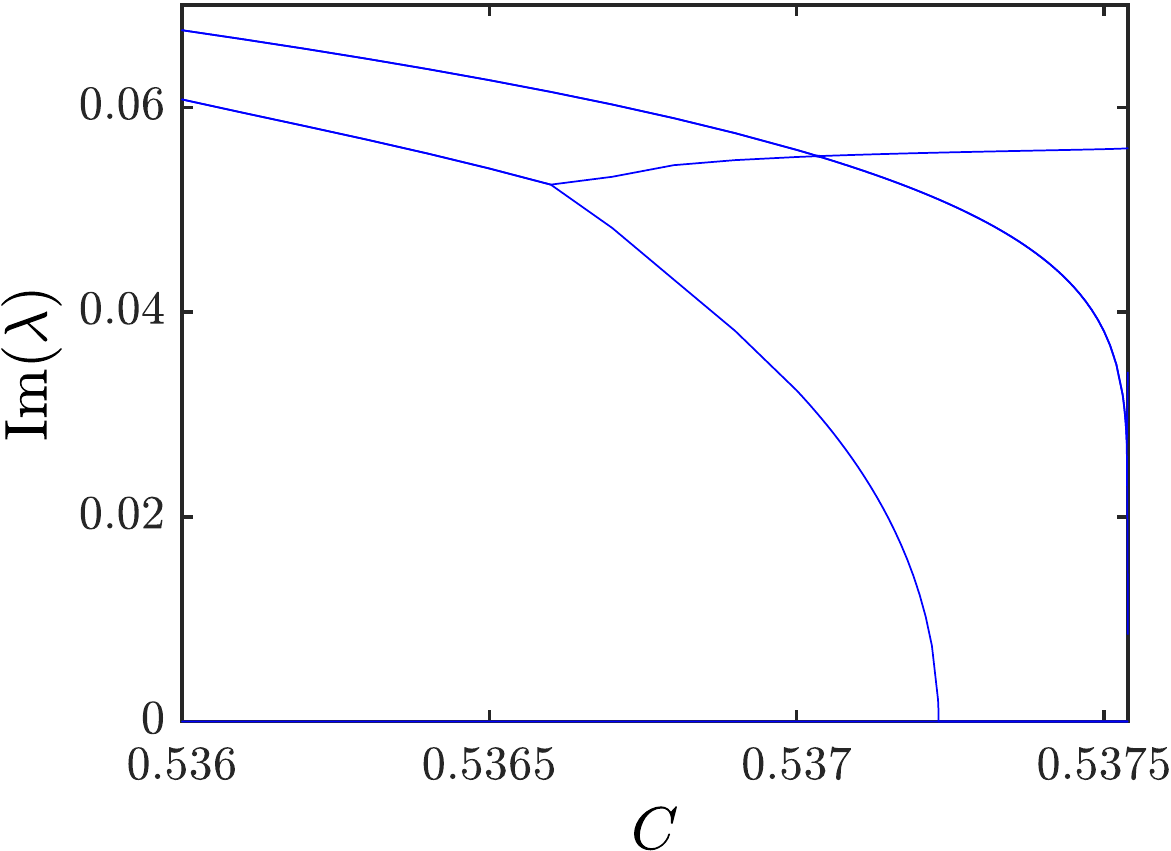} &
\includegraphics[width=6cm]{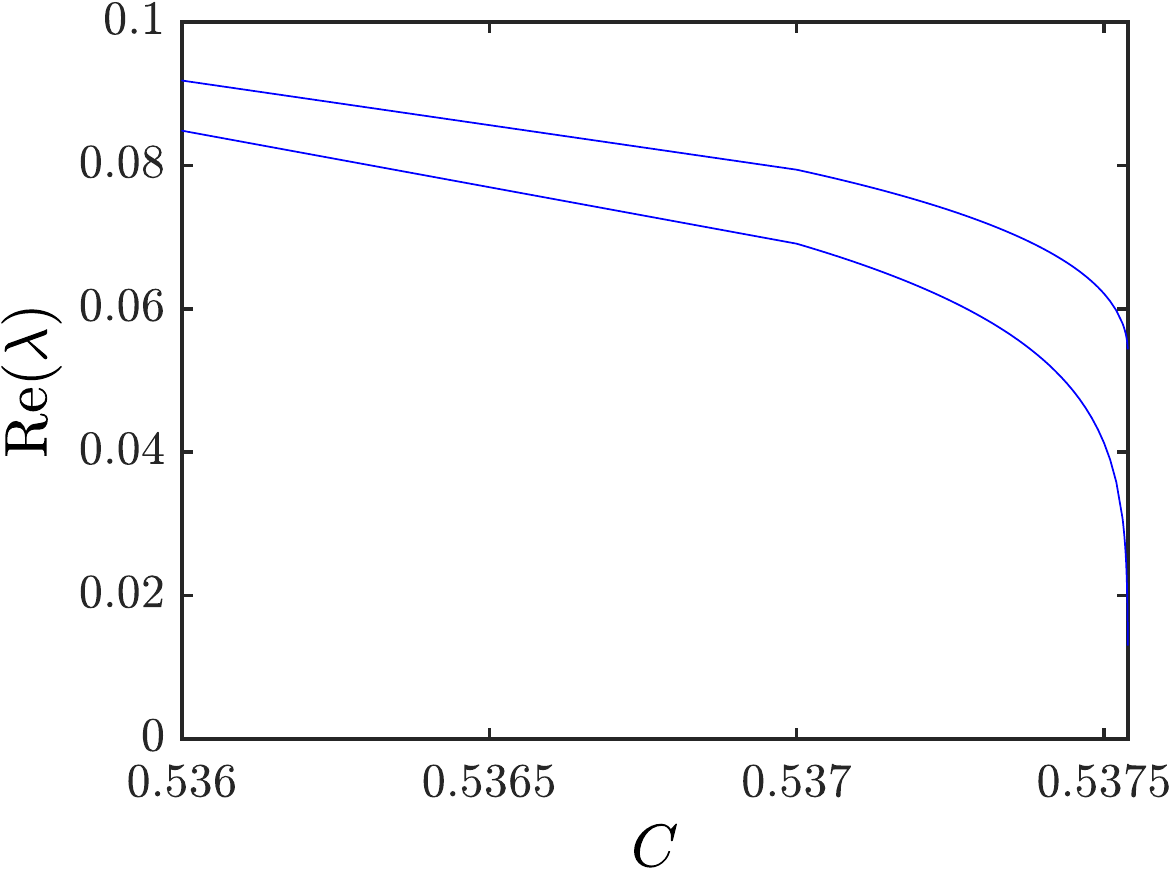} \\
\end{tabular}
\caption{(left) Imaginary part of the eigenvalues of counterwinding 2VSs, and (right) real part of the eigenvalues of counterwinding 0VSs, close to the pitchfork and turning points. The collision of the two branches occurs at $C_t$, where they disappear in a turning point bifurcation.}
\label{fig:stab2}
\end{figure}

Figure~\ref{fig:stab1} depicts the dependence on $C$ of the real and
imaginary part of the eigenvalues for counterwinding 2VSs, 1VS, and
0VSs. In these images we have used a $41\times 41$ lattice, but as
previously remarked, the results remain nearly identical
for different lattice sizes. We can see that the 1VS and 0VSs are exponentially unstable for every $C\geq 0$ for which they respectively exist since they both have eigenvalues $\lambda$ whose real parts are only vanishing at their bifurcation points $C_p$ and $C_t$, respectively. In Figure~\ref{fig:stab2} we provide a zoomed in version of Figure~\ref{fig:stab1} close to $C_p$ and $C_t$ for the imaginary part of the eigenvalues of 2VSs and the real part of the eigenvalues of 0VSs, respectively. Here we find that one of the NEE pairs arrives at $\lambda = 0$ at $C = C_p$ so that the vortex pair becomes exponentially unstable past the bifurcation point due to its subcritical pitchfork bifurcation with the 1VS solution branch. The continuation past this critical point of the subsequently exponentially unstable 2VS solution branch stops when the remaining NEE pair reaches $\lambda = 0$ at $C = C_t$. There, the collision occurs with the 0VS solution branch and the termination of both of these branches takes place. No stationary solutions involving two counterwinding vortices at such a distance can be identified past this critical point.
Both (the 0VS and the 2VS) branches have a positive real eigenvalue pair and a zero eigenvalue pair at the bifurcation point of $C=C_t$.
Notice that supposing the previous linear dependence of the NEEs $\lambda=\pm\mathrm{i}(1-2C)$ leads to a zero eigenvalue at $C=0.5$, a value which is very close to the bifurcation point observed in Figure~\ref{fig:bifsN}.

The case of cowinding vortices is similar. Moving away from the anti-continuum limit there exists a $C_c \in (0,C_p)$ for which the NEEs of the cowinding 2VS collide with the continuous band, creating a subsequent cascade of Hamiltonian Hopf and inverse Hamiltonian Hopf bifurcations. This gives an oscillatory instability for the cowinding 2VS state for $C$ above $C_c$. We note that for $C<C_c$ the linear dependence of the NEEs is the same as in the counterwinding case, and, consequently, the value of $C_c$ is also the same for both cases.

The major difference between the counterwinding and the cowinding spectra concerns the location of the pitchfork bifurcation (and the branches it
involves). Recall that in the counterwinding case the pitchfork bifurcation takes place on the 2VS bifurcation curve, whereas in the cowinding case the pitchfork bifurcation takes place on the 0VS bifurcation curve. Hence, the NEEs of the 2VSs in the cowinding case do not arrive at $\lambda = 0$ prior to reaching $C = C_t$. Therefore the cowinding 2VSs only exhibit oscillatory instabilities as the turning point is approached. At the turning point, one of the NEE pairs becomes zero. On the other hand, it is the 1VS branch that collides in the supercritical pitchfork bifurcation with the 0VS branch. The latter possesses 2 real eigenvalue pairs in the vicinity of the anti-continuum limit, while it only has 1 such in the vicinity of the
turning point (past the pitchfork bifurcation). The 1VS branch, as in the counterwinding case, carries one real eigenvalue pair and a potential additional oscillatory instability due to an NEE mode.

\subsection{Dynamic Solutions}\label{subsec:Dynamics} 

For the convenience of the reader, this subsection is broken down into two distinct components - one for counterwinding vortices and one for cowinding vortices.

\subsubsection{Counterwinding Vortices}

In the continuum limit it is known that counterwinding vortices move in parallel along a straight line, whereas here numerical investigations will reveal that this is not the case in the context of our lattice dynamical system (\ref{dNLS}). Here we will focus on the instabilities of the 2VS solutions for $C \in (C_c,C_t)$ to explore the expected dynamics for solutions that start very close to our stationary 2VSs. All images of simulations in this section are taken for an $81\times81$ lattice in an effort to minimize the boundary effects on the dynamics.

\begin{figure}
\begin{center}
\includegraphics[width=12cm,clip=true]{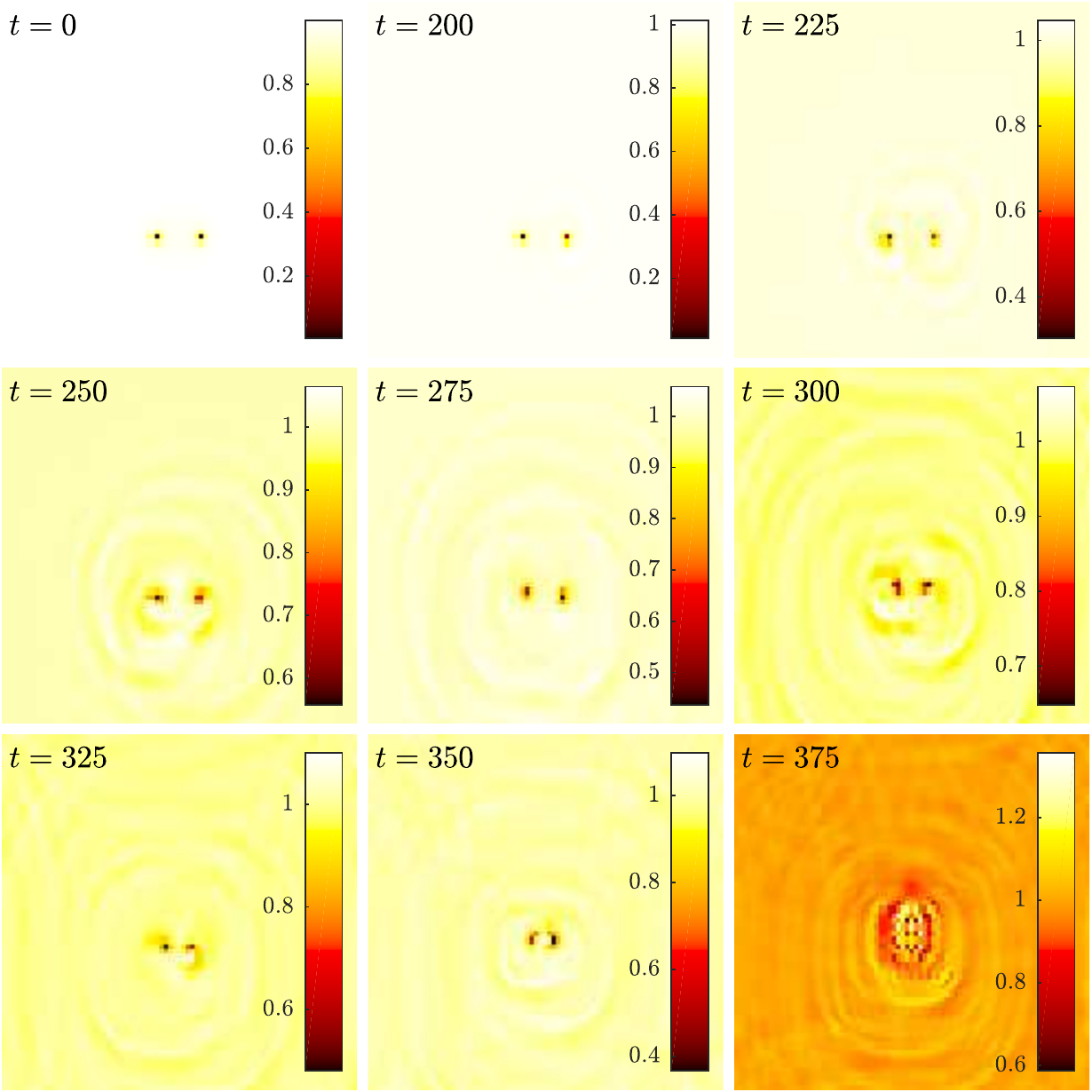}
\end{center}
\caption{Dynamics of (\ref{dNLS}) with a randomly perturbed initial
  condition near the stationary counterwinding 2VS at $C=0.25$. At
  this parameter value the 2VS pair only possesses oscillatory
  instabilities. As can be seen, the vortices translate, yet also
  approach each other and eventually pair-wise annihilate. {The panels depict the density $|\psi_{n,m}|^2$ at selected values of time.}}
\label{fig:dyncounter1}
\end{figure}

We first begin by adding a small random perturbation ($\sim 10^{-8}$) to 2VSs with $C$ in the interval $(C_c,C_t)$. We have found that all simulations lead to upward (i.e., perpendicular to the axis connecting the two vortices) translating vortices which eventually collide at some finite time step. This can be observed in Figure~\ref{fig:dyncounter1} where we take $C = 0.25$ and provide a number of snapshots of the dynamic evolution. Notice that the counterwinding vortices appear to be propagating upward with a slight bend toward each other, leading {at $t\approx360$} for this particular simulation
to their collision and pair annihilation. We note that our random perturbation does have the effect of breaking the $\psi_{n,m} = \psi_{-n,m}$ symmetry of the counterwinding vortices, and therefore we have also explored initial conditions which preserved this symmetry. That is, if our solution is given by $\psi_{n,m} = r_{n,m}\mathrm{e}^{\mathrm{i}\theta_{n,m}}$, then we introduce the initial condition $\psi_{n,m}(0) = (r_{n,m}+\delta)\mathrm{e}^{\mathrm{i}\theta_{n,m}}$ so that $\psi_{n,m}(0) = \psi_{-n,m}(0)$. In this case the evolution in $t$ preserves the $(n,m)\mapsto(-n,m)$ symmetry of the initial condition, but again we find that the vortices eventually collide and annihilate each other.
Moreover, this dynamical outcome arises both for the case of oscillatory
instabilities (as in Fig.~\ref{fig:dyncounter1}), and for that
of exponential instabilities (not shown here).

For values of the coupling constant $C$ taken beyond $C_t$ we no longer have stationary counterwinding vortex solutions. We do however take the stationary solutions near $C = C_t$ and use them as initial conditions for coupling values well beyond the turning point. The aim of this is to explore the behavior of the vortex
pair as the continuum limit is approached. Interestingly our temporal evolution reveals that the counterwinding vortices continue to propagate upward and eventually collide and annihilate each other. We further find that as the coupling constant $C$ grows larger, it takes a longer time for the vortices to approach each other and
collide. In \S~\ref{sec:Continuum} we will present formal arguments which appear to explain some of this behaviour. Our observations (and formal arguments to follow) thus suggest that the genuine traveling of counterwinding vortices at the continuum limit is a singular behavior that is ``destroyed'' by discreteness, rather than a behavior that potentially bifurcates at some finite value of $C$. Instead, discreteness introduces a (weaker, the larger the coupling strength) lateral dynamical motion of the vortex pair, leading eventually to its apparently generic for finite $C$ annihilation.

\subsubsection{Cowinding Vortices}

\begin{figure}
\includegraphics[width=12cm,clip=true]{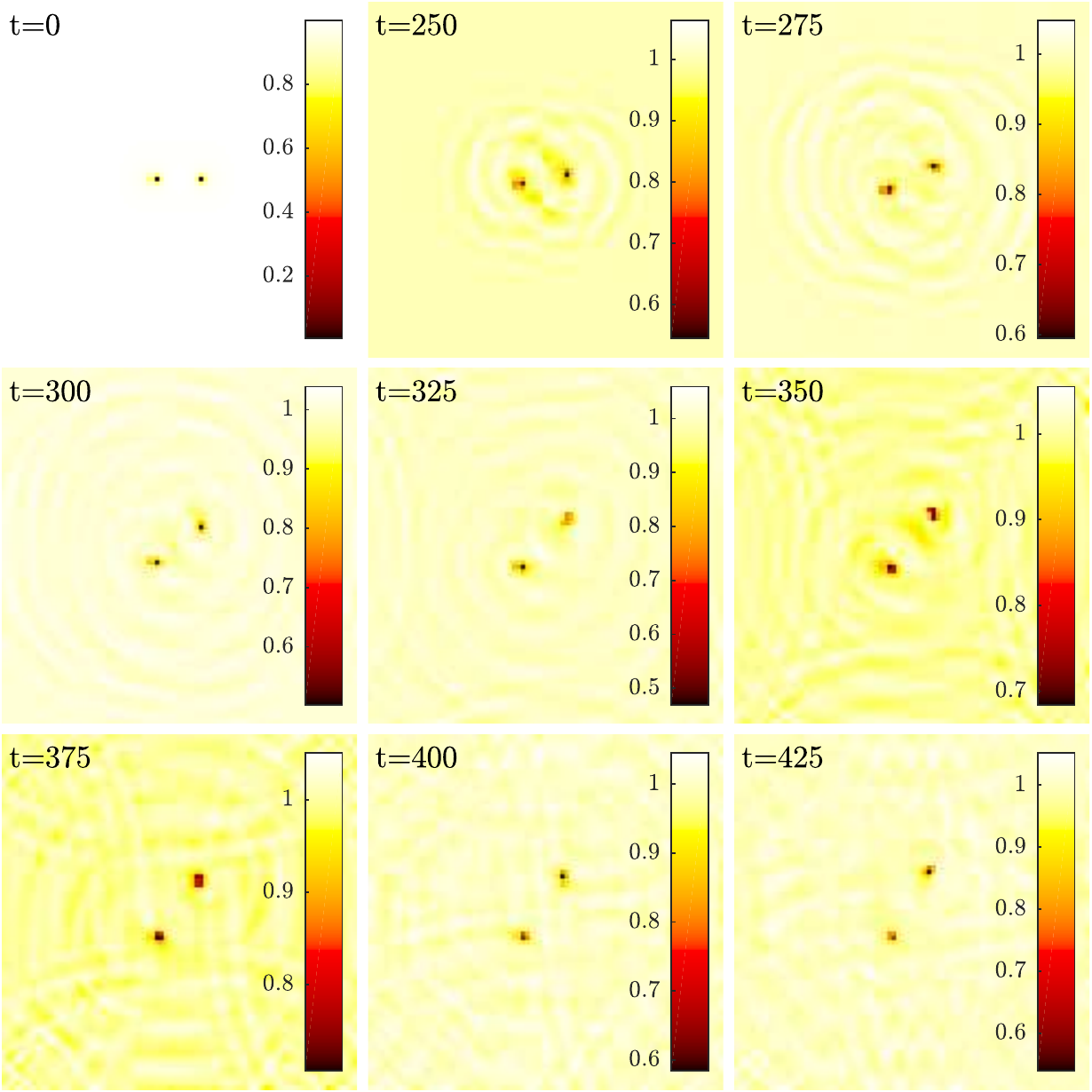} \\
\caption{Dynamics of (\ref{dNLS}) with a randomly perturbed initial condition near the stationary cowinding 2VS at $C=0.23$. The dynamical evolution
leads to rotation around the $(n,m) = (0,0)$ lattice site, but then begin to slow down and appear to halt resulting in a rotated
(and with larger distance between the vortices than the original one)
vortex configuration. {The panels depict the density $|\psi_{n,m}|^2$ at selected values of time.}}
\label{fig:dyncow1}
\end{figure}

We now numerically observe the dynamical evolution of solutions that start near our cowinding vortices. We recall
that, as discussed above, the cowinding 2VSs only experience oscillatory instabilities. As a result of these, the waveforms which are initialized
near these stationary solutions start to rotate about the $(n,m) = (0,0)$ lattice site but eventually slow down their rotation and appear to stop, resulting in a ``pseudo-stationary'' cowinding vortex configuration. Nevertheless, notice that it is less straightforward to extract the asymptotic scenario in this case in part due to the residual radiation present in the dynamical lattice.
A relevant evolution is exemplified in Figure~\ref{fig:dyncow1} where we provide snapshots of the time stepping with initial condition given by a random perturbation of a 2VS state at $C = 0.23$. It is important to notice, however, that the dynamically resulting stationary configuration involves vortices at a larger distance than the initial one.

We also report that the exponential instabilities of the 0VSs appear to lead to vortices which rigidly rotate about the $(n,m) = (0,0)$ lattice for all $t\geq 0$. It is important to mention that such rotation does not happen at constant inter-center separation between the vortices. Rather, as we see in more detail also below, the distance between the cowinding pair member vortices increases over time in a spiraling out motion (see below for a demonstration of such a spiralling dynamical example).

\begin{figure}
\includegraphics[width=12cm,clip=true]{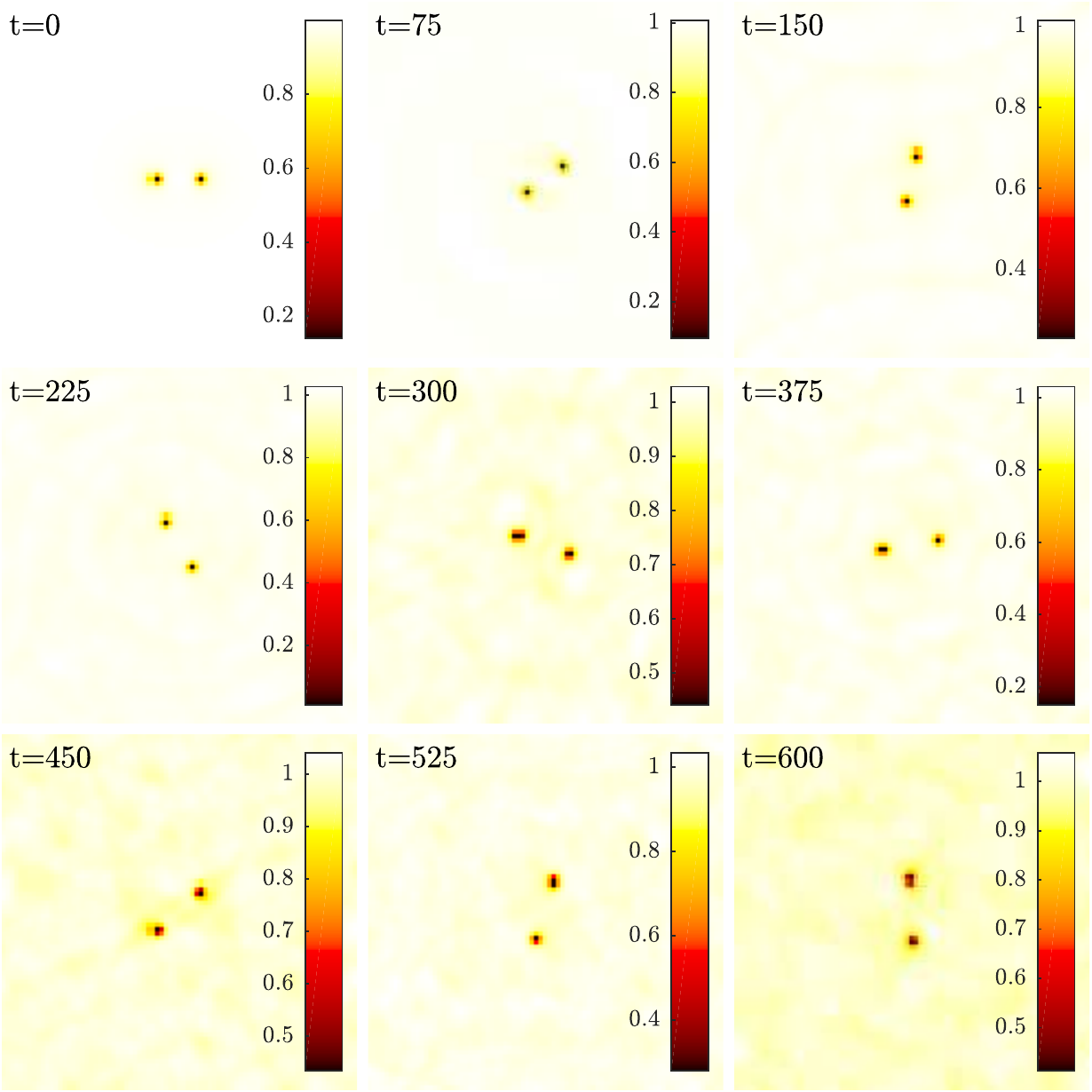} \\
\caption{Dynamics of (\ref{dNLS}) with a stationary cowinding 2VSs at $C = 0.498$ used as an initial condition with $C=0.51$. Note that at $C = 0.51$ we have no longer stationary cowinding vortices and now we can see that the vortices begin to rotate around each other, spiralling outward
(i.e., with a growing distance between them) over time. {The panels depict the density $|\psi_{n,m}|^2$ at selected values of time.}}
\label{fig:dyncow2}
\end{figure}

Stationary cowinding vortex solutions no longer exist for values of the coupling constant $C$ taken beyond $C_t$, and therefore we will explore what happens when a stationary 2VS state taken at some $C$ slightly below $C_t$ is used as an initial condition in (\ref{dNLS}) for coupling
$C > C_t$. Our temporal evolution demonstrates that these cowinding vortices hold their shape but begin to rigidly rotate about the lattice site $(n,m) = (0,0)$, as is demonstrated in Figure~\ref{fig:dyncow2}. This type of temporal dynamics resembles the corresponding evolution
in the continuum limit ($C\to \infty$) of (\ref{dNLS}), however with a significant modification. In particular, over time the vortices rotate
at larger distances from each other and do so more slowly (i.e., at smaller angular momentum). This is the by-product of discreteness
once again presumably destroying the perfectly rotating continuum states, in favor of progressively separating (non-periodic) discrete ones.
As the continuum limit is approached, this lateral motion is still present although it
becomes weaker, suggesting that it only disappears in the (singular) continuum limit.

\section{The Continuum Equation}\label{sec:Continuum} 

In this section we provide formal arguments in an attempt to describe some of the expected counter- and cowinding vortex dynamics in (\ref{dNLS}) for $\varepsilon$ (or equivalently $C$) large.
More specifically, our aim here is to illustrate the plausibility of the non-existence of rigidly translating (for counterwinding) or
rotating (for cowinding) vortex states in the genuinely discrete DNLS problem.
We begin by noting that taking $\varepsilon \to \infty$ in (\ref{dNLS}) marks a return to the well-studied non-linear Schr\"odinger equation in two continuous spatial dimensions given by~\cite{siambook}
\begin{equation}\label{NLS}
	\mathrm{i}\frac{\partial\psi}{\partial t} - |\psi|^2\psi + \frac{\partial^2\psi}{\partial x^2} + \frac{\partial^2\psi}{\partial y^2}= 0, \quad (x,y) \in \mathbb{R}^2,
\end{equation}
where $\psi = \psi(x,y,t)$ is a complex-valued function. Equation (\ref{NLS}) possesses an important symmetry property: if $\psi(x,y,t)$ is a solution to (\ref{NLS}) then so is
\[
	\tilde{\psi}(x,y,t) = \psi(\cos(\theta)(x+p_1) - \sin(\theta)(y + p_2),\sin(\theta)(x + p_1) + \cos(\theta)(y + p_2),t),
\]
for any angle $\theta \in S^1$ and translation $(p_1,p_2)\in\mathbb{R}^2$. These rotations and translations together form the special Euclidean group, often denoted as ${\bf SE}(2)$, and equation (\ref{NLS}) precisely is said to be invariant with respect to the action of this group. Given a function $\psi:\mathbb{R}^2 \to \mathbb{C}$, the {\em group orbit} of $\psi$ is given by the set
\[
	\begin{split}
	{\bf SE}(2)\psi := \{&\psi(\cos(\theta)(x+p_1) - \sin(\theta)(y + p_2),\sin(\theta)(x + p_1) + \cos(\theta)(y + p_2)): \\ &\quad \theta\in S^1,\ (p_1,p_2)^T\in\mathbb{R}^2\},
	\end{split}
\]
which is simply the application of every element of ${\bf SE}(2)$ to the function $\psi$. Then, a group orbit $X$ is a {\em relative equilibrium} if the flow of (\ref{NLS}) leaves $X$ invariant. That is, relative equilibria of (\ref{NLS}) are equilibrium solutions in a moving coordinate frame. A trivial example of a relative equilibrium would be any equilibrium of (\ref{NLS}) since any function trivially belongs to its own group orbit.

Counterwinding vortex solutions of (\ref{NLS}) linearly propagate with constant nonzero speed~\cite{siambook}, meaning that their temporal evolution is described by a continuous linear translation and therefore they belong to their group orbit for all $t \in \mathbb{R}$. Similarly, cowinding vortex solutions of (\ref{NLS}) rotationally propagate with constant nonzero angular frequency, meaning that their temporal evolution is described by a continuous rotation and therefore belong to their group orbit for all $t \in \mathbb{R}$. Hence, cowinding and counterwinding vortex solutions are relative equilibria of (\ref{NLS}) and in turn their respective group orbits define invariant manifolds in some appropriate complex-valued space of functions with domain $\mathbb{R}^2$. Most importantly, these invariant manifolds are exactly three-dimensional, corresponding to the three degrees of freedom of the special Euclidean group {\bf SE}(2) (i.e. two dimensions of translations and one dimension of rotations). Much work has been undertaken to capture the qualitative dynamics of relative equilibrium solutions using the group orbit, particularly to understand
a closely related example, namely the dynamics of spiral waves as solutions to reaction-diffusion equations \cite{Ashwin,Victor3,SSW}. We will use some of these theoretical results in an attempt to understand our observations for (\ref{dNLS}) far from the anti-continuous limit.

An effective way to understand the dynamics of (\ref{dNLS}) with large $\varepsilon$ is to apply an inhomogeneous symmetry breaking perturbation to (\ref{NLS}) which preserves only the symmetries of a square lattice \cite{Victor1,Victor2}. Such symmetry-breaking perturbations can be used to mimic the effect of discretizing space by breaking the continuous translational and rotational symmetries of the two-dimensional non-linear Schr\"odinger equation (\ref{NLS}). Hence, let us consider a small parameter $0 \leq \delta \ll 1$ and some sufficiently smooth function $\mathcal{F}$ so that we perturb (\ref{NLS}) as
\begin{equation}\label{NLS_Pert}
	\mathrm{i}\frac{\partial\psi}{\partial t} - |\psi|^2\psi + \frac{\partial^2\psi}{\partial x^2} + \frac{\partial^2\psi}{\partial y^2} + \delta\mathcal{F}(x,y,\psi,\delta)= 0.
\end{equation}
The specific form of $\mathcal{F}$ is not important and can be generalized, but here we will assume that it is $1$-periodic and even in both $x$ and $y$, and linear in $\psi$. The linearity of $\mathcal{F}$ with respect to $\psi$ allows one to eliminate the oscillatory component $\mathrm{e}^{-\mathrm{i}\omega t}$ and still obtain an autonomous equation, as was done for the discrete non-linear Schr\"odinger equation (\ref{dNLS}) in the previous sections. One should note that an important, and possibly motivating, example of such a function $\mathcal{F}$ would be
\[
	\mathcal{F}(x,y,\psi,\delta) = V(x,y)\psi,	
\]
where $V(x,y)$ is a potential that respects the symmetries of the two-dimensional integer lattice. Therefore, our goal is to hypothesize how the dynamics of the invariant manifolds for the unperturbed Schr\"odinger equation (\ref{NLS}) given by the group orbits will perturb for $0 < \delta \ll 1$. This should inform us about what to expect
regarding the behaviour of the cowinding and counterwinding vortices in the discrete spatial context of (\ref{dNLS}).

\subsection{Counterwinding Vortices} 

We begin by focusing on the case of counterwinding vortices. Such vortex solutions to (\ref{NLS}) can be found by using the ansatz
\begin{equation}\label{CounterAnsatz}
	\psi(x,y,t) = A(x,y-vt)\mathrm{e}^{-\mathrm{i}\omega t},
\end{equation}
with $\xi = y-vt$, and constant $v \in \mathbb{R}$. We refer to $A$ as the profile of the counterwinding vortex pair, and we note these counterwinding vortices of (\ref{NLS}) are partially characterized by their $x \mapsto -x$ symmetry, implying that the profile $A$ is even in $x$. The set of all functions which are even in $x$ is flow-invariant for (\ref{NLS}), and hence an ansatz of the form (\ref{CounterAnsatz}) represents the intersection of this flow-invariant subspace of even functions and the group orbit of counterwinding vortices. The profile $A$ can be obtained by solving
\begin{equation}\label{NLS_Counter}
	 A_{xx} + A_{\xi\xi} + (\omega - |A|^2)A - \mathrm{i}vA_{\xi} = 0,	
\end{equation}
where the subscripts denote partial differentiation. We assume that for some fixed $\omega > 0$, there exists a real $\bar{v} \neq 0$ such that a counterwinding vortex profile $A_0$ is a solution of (\ref{NLS_Counter}) with $v = \bar{v}$. That is,
\[
	\psi(x,y,t) = A_0(x,y-\bar{v}t)\mathrm{e}^{-\mathrm{i}\omega t}
\]
is a counterwinding vortex pair solution of (\ref{NLS}) and that $A(-x,y-vt) = A(x,y-vt)$ for all $(x,y,t)$.  The ansatz (\ref{CounterAnsatz}) implies that our solution is propagating linearly parallel to the $y$-axis, but recall that the ${\bf SE}(2)$ invariance of (\ref{NLS}) implies that we could make it propagate in any direction we want.

Now equation (\ref{NLS_Pert}) is a non-autonomous perturbation of (\ref{NLS}), which, in turn, yields that the flow along the perturbed invariant manifold coming from the group orbit of counterwinding vortices will also be non-autonomous. Moreover, since the set of all functions which are even in $x$ is again flow-invariant for (\ref{NLS_Pert}), we again restrict ourselves to the intersection of this flow-invariant subspace and the perturbed invariant manifold. Let us assume that the function $A_\delta(x,y)$ is the profile of a counterwinding vortex solution to (\ref{NLS_Pert}) for $0 \leq \delta \ll 1$ which is even in $x$. Then, the results of \cite{Victor1,Victor2} lead one to believe that the continued counterwinding vortex solution of (\ref{NLS_Pert}) for small $\delta > 0$ is of the form
\begin{equation}\label{NLS_Counter2}
	\psi(x,y,t) = A_\delta(\cos(\alpha(x,y,\delta))\xi_1 - \sin(\alpha(x,y,\delta))\xi_2,\sin(\alpha(x,y,\delta))\xi_1 + \cos(\alpha(x,y,\delta))\xi_2)\mathrm{e}^{-\mathrm{i}\omega t},	
\end{equation}
where $\xi_1 = x - v_x(x,y,\delta)t$ and $\xi_2 = y - v_y(x,y,\delta)t$. Moreover, the functions $\alpha,v_x,$ and $v_y$ are uniformly bounded, $\alpha$ is $1$-periodic in both $x$ and $y$, and satisfy
\[
	\alpha(x,y,\delta) = \mathcal{O}(\delta), \quad v_x(x,y,\delta) = \mathcal{O}(\delta), \quad v_y(x,y,\delta) = \bar{v} + \mathcal{O}(\delta), 	
\]
where $\bar{v}$ is the speed of the counterwinding vortex from the unperturbed equation (\ref{NLS}). One should interpret the function $\alpha(x,y,\delta)$ as introducing a slight wobble into the motion of the vortex solution, whereas $v_x(x,y,\delta)$ and $v_y(x,y,\delta)$ describe the inhomogeneous speed of linear propagation in the $x$ and $y$ directions, respectively.

From the form of (\ref{NLS_Counter2}), we have that for any fixed $(a_1,a_2)\in\mathbb{R}^2$, the function $A_\delta$ is constant along the (generically one-dimensional) level sets
\[
	\begin{split}
		\cos(\alpha(x,y,\delta))(x - v_x(x,y,\delta)t) - \sin(\alpha(x,y,\delta))(y - v_y(x,y,\delta)t) &= a_1, \\	
		\sin(\alpha(x,y,\delta))(x - v_x(x,y,\delta)t) + \cos(\alpha(x,y,\delta))(y - v_y(x,y,\delta)t) &= a_2,
	\end{split}
\]
for each fixed $\delta > 0$ sufficiently small. These curves serve as
characteristics for the counterwinding vortex solution to the
inhomogeneous equation (\ref{NLS_Pert}). Requiring that the solution
(\ref{NLS_Counter2}) is even in $x$ gives that $\alpha$ and $v_y$ are
both even in $x$ and $v_x$ is odd in $x$, and these symmetries imply
that for every $a_1 \neq 0$ and $a_2\in\mathbb{R}$, the characteristic
curve corresponding to the level set $(a_1,a_2)$ and the
characteristic curve corresponding to the level set $(-a_1,a_2)$ are
continuously mapping into each other by the action
$(x,y,t)\mapsto(-x,y,t)$. Furthermore, in the case when
$v_x(x,y,\delta) \neq 0$ for all $x \neq 0$, $y\in\mathbb{R}$ and $0 <
\delta \ll 1$, we have that these characteristic curves approach $x =
0$ in either forward or backward $t$. In this case we would observe an
$\mathcal{O}(\delta)$ drift in the $x$-direction of the solution to
the middle in either forward or backward time. Hence, these heuristic
arguments seem to suggest that it is exactly the $n \to -n$ symmetry
(the discrete analogue of $x \to -x$ symmetry) of the solutions to the
discrete non-linear Schr\"odinger equation which drives the breakup of
solutions we observed previously in the dynamics.
Indeed, it is this drift of the vortex centers towards $x=0$ which leads to their pairwise approach and eventual annihilation
observed in the dynamics of Fig.~\ref{fig:dyncounter1}.

\subsection{Cowinding Vortices} 

We now initiate a similar exploration for cowinding vortex solutions to show that the $(n,m)\to(-n,-m)$ symmetry of cowinding vortices is expected to drive the breakup observed in simulations of (\ref{dNLS}) far from the anti-continuous limit. These vortex solutions to (\ref{NLS}) can be found by using the ansatz
\begin{equation}\label{CoAnsatz}
	\psi(x,y,t) = B(\cos(\beta t)x - \sin(\beta t)y,\sin(\beta t)x + \cos(\beta t)y)\mathrm{e}^{-\mathrm{i}\omega t},
\end{equation}
for constants $\beta \neq 0$ and $\omega \geq 0$. We refer to $B$ as the profile of the cowinding vortex pair, and we note that the cowinding vortices of (\ref{NLS}) are partially characterized by their $(x,y)\mapsto (-x,-y)$ symmetry, which is imposed by requiring that $B(-x,-y) = B(x,y)$. The set of all functions which are invariant with respect to $(x,y)\mapsto (-x,-y)$ symmetry is flow-invariant for (\ref{NLS}), and hence an ansatz of the form (\ref{CounterAnsatz}) represents the intersection of this flow-invariant subspace and the group orbit of cowinding vortices. We note that now cowinding vortices are rigidly rotating in space with angular velocity $\beta$, implying that the characteristic curves of solutions of the form (\ref{CoAnsatz}) are closed concentric circles about the origin $(x,y) = (0,0)$, as opposed to straight lines in the counterwinding case. We assume that for some fixed $\omega > 0$ there exists a real $\beta^* \neq 0$ and a profile $B_0$ so that
\[
	\psi(x,y,t) = B_0(\cos(\beta^* t)x - \sin(\beta^* t)y,\sin(\beta^* t)x + \cos(\beta^* t)y)\mathrm{e}^{-\mathrm{i}\omega t},	
\]
is a cowinding vortex solution of (\ref{NLS}) satisfying $B_0(-x,-y) = B_0(x,y)$ for all $(x,y) \in \mathbb{R}^2$.

As previously remarked, the equation (\ref{NLS_Pert}) is a non-autonomous perturbation of (\ref{NLS}), and hence the perturbed invariant manifold coming from the group orbit of counterwinding vortices will also be non-autonomous. Moreover, since the set of all functions which are invariant with respect to $(x,y)\mapsto (-x,-y)$ symmetry is again flow-invariant for (\ref{NLS_Pert}), we again restrict ourselves to the intersection of this flow-invariant subspace and the perturbed invariant manifold. Then, let us assume that the function $B_\delta(x,y)$ is the profile of a cowinding vortex solution to (\ref{NLS_Pert}) for $0 \leq \delta \ll 1$, which satisfies $B_\delta(-x,-y) = B_\delta(x,y)$. This leads one to conjecture that the continued cowinding vortex solution of (\ref{NLS_Pert}) for small $\delta > 0$ is of the form
\begin{equation}\label{NLS_Co2}
	\psi(x,y,t) = B_\delta(\cos(\beta(x,y,\delta))\zeta_1 - \sin(\beta(x,y,\delta))\zeta_2,\sin(\beta(x,y,\delta))\zeta_1 + \cos(\beta(x,y,\delta))\zeta_2)\mathrm{e}^{-\mathrm{i}\omega t},	
\end{equation}
where $\zeta_1 = x - d_x(x,y,\delta)t$ and $\zeta_2 = y - d_y(x,y,\delta)t$. The functions $\beta,d_x,$ and $d_y$ are uniformly bounded and satisfy
\[
	\beta(x,y,\delta) = \beta^* + \mathcal{O}(\delta), \quad d_x(x,y,\delta) = \mathcal{O}(\delta), \quad d_y(x,y,\delta) = \mathcal{O}(\delta), 	
\]
where $\beta^*$ is the rotational velocity of the unperturbed cowinding vortex. Hence, we see that we should expect $\mathcal{O}(\delta)$ drifts in both the $x$- and $y$-directions.

From the form of (\ref{NLS_Co2}), we have that for any fixed $(b_1,b_2)\in\mathbb{R}^2$, the function $B_\delta$ is constant along the (generically one-dimensional) level sets
\[
	\begin{split}
		\cos(\beta(x,y,\delta))(x - d_x(x,y,\delta)t) - \sin(\beta(x,y,\delta))(y - d_y(x,y,\delta)t) &= b_1, \\	
		\sin(\beta(x,y,\delta))(x - d_x(x,y,\delta)t) + \cos(\beta(x,y,\delta))(y - d_y(x,y,\delta)t) &= b_2,
	\end{split}
\]
for each fixed $\delta > 0$ sufficiently small. These curves serve as characteristics for the cowinding vortex solution to the inhomogeneous equation (\ref{NLS_Pert}). Requiring that the solution (\ref{NLS_Co2}) satisfies $\psi(-x,-y,t) = \psi(x,y,t)$ requires that
\[
	\beta(-x,-y,\delta) = \beta(x,y,\delta), \quad d_x(-x,-y,\delta) = -d_x(x,y,\delta), \quad d_y(-x,-y,\delta) = -d_y(x,y,\delta),
\]
for all $(x,y)\in\mathbb{R}^2$ and sufficiently small $\delta \geq 0$. Note that this implies that $d_x(0,0,\delta) = d_y(0,0,\delta) = 0$. Hence, these symmetries imply that for every $(b_1,b_2) \neq (0,0)$, the characteristic curve corresponding to the level set $(b_1,b_2)$ and the characteristic curve corresponding to the level set $(-b_1,-b_2)$ are continuously mapping into each other by the action $(x,y,t)\mapsto(-x,-y,t)$. Furthermore, in the case when $d_x(x,y,\delta)\cdot d_y(x,y,\delta) > 0$ for all $(x,y)\in\mathbb{R}\setminus\{(0,0)\}$ and $0 < \delta \ll 1$, we have that these characteristic curves approach $(x,y) = (0,0)$ in either forward or backward $t$. In this case we would observe an $\mathcal{O}(\delta)$ spiral into to the centre $(x,y) = (0,0)$ in either forward or backward time. Hence, as before, we have provided heuristic arguments detailing the behaviour of (\ref{dNLS}) near the continuum limit, i.e. $\varepsilon \gg 0$.
Indeed, this is also in line with our observations namely the finding that over (positive) time, the distance between the
vortices slightly increases, i.e., that they are spiraling outwards from the center. As they do so, once again a rigidly rotating
configuration cannot be reached; cf. Fig.~\ref{fig:dyncow1}, except
at the continuum limit.

\section{Conclusions and Future Challenges}

In the present work, we have extended considerations associated with a single vortex in a discrete nonlinear
Schr{\"o}dinger setting to ones involving vortex pairs of either the same or of opposite charges.
These configurations in the continuum limit of the equation either rotate rigidly (for same charge)
or translate with constant speed (for opposite charges). These tendencies are contrasted with
the limit of vanishing coupling, the so-called anti-continuum limit which halts the vortex motion. This raises
the interesting question of what happens ``in between'', i.e.,
for coupling strengths between $C \rightarrow 0^+$ and
$C \rightarrow \infty$. We find that in the vicinity of the vanishing coupling, stationary configurations
can exist involving the two vortices (of either same or of opposite charge). These configurations
present an interesting sequence of bifurcation phenomena involving pitchfork (symmetry breaking)
bifurcations, as well as saddle-center, turning-point ones. The latter lead to the termination
of the stationary multi-vortex branches, raising once again the question of how the continuum limit
group orbit (rotation or translation) motions arise. Indeed, the answer to this question is non-trivial
too. We find that vortex pairs initialized past these turning points cannot rigidly rotate or steadily
translate. Rather, in the cowinding case instead of rotating, they spiral out. In the counterwinding
one, rather than translate, they approach each other (while moving) and eventually annihilate.
The rate of these lateral motions appears to decrease as the coupling increases, and it seems reasonable
to conjecture that these motions only disappear altogether in the singular continuum limit.

Our numerical computations, we believe, shed some light on the system's phenomenology. However, admittedly,
they also raise numerous interesting questions for future investigation both at the mathematical and
at the computational, as well as at the physical level. More specifically, quantifying the rate of
spiraling for the cowinding case, and that of lateral approach in the counterwinding one (as a function
of the coupling strength
$C$) is an important question for future analytical
and numerical consideration. On the other hand, proving rigorously
the non-existence of discrete rotation or translation for finite $C$ is of interest in its own right.
Providing a rigorous characterization of the stability of the 0VS, 1VS and 2VS states is also an interesting
task from the opposite, near-anti-continuum limit. Naturally, all of these considerations are worthwhile
to extend in the context of three-dimensional systems. There, it is well-known that configurations
such as vortex lines and vortex rings represent the
principal topologically charged entities in the dynamics~\cite{siambook}.
Presently, we are not aware of any studies exploring systematically the stability of such states
as regards either a single structure or pairs thereof. Such a study would be particularly
interesting because, e.g., for vortex rings even a single one is subject to translation in the
continuum limit~\cite{siambook}. Hence it is relevant to explore
the impact of discreteness near the $C=0$ limit. Moreover, remarkable phenomena such as leapfrogging dynamics arise, e.g., as a result
of the interaction of multiple vortex rings~\cite{caplan}, hence it is natural to inquire about their
fate in the discrete realm. Such studies are currently in progress and will be reported in future publications.

\section*{Acknowledgements}
J.J.B. was supported by an NSERC PDF. This material is based upon work supported by the National Science Foundation under Grant No. PHY-1602994 and
under Grant No. DMS-1809074 (P.G.K.). P.G.K. also acknowledges support from QNRF via the program NPRP-9-329-1-067. J.C.-M. thanks financial support from MAT2016-79866-R project (AEI/FEDER, UE).



\begin{thebibliography}{99}

\bibitem{mark2} M.J. Ablowitz and J.T. Cole, Phys. Rev. A {\bf 96}, 043868 (2017).

\bibitem{Ashwin} P. Ashwin, I. Melbourne, Nonlinearity {\bf 10}, 595 (1997).

\bibitem{Bramburger} J.J. Bramburger, J. Dyn. Differ. Equ. {\bf 31}, 469 (2019).

\bibitem{caplan} R.M. Caplan, J.D. Talley, R. Carretero-Gonz{\'a}lez, P.G. Kevrekidis, Phys. Fluids {\bf 26}, 097101 (2014).

\bibitem{Victor1} L. Charette, V.G. LeBlanc, SIAM J. Appl. Dyn. Syst. {\bf 13}, 1694 (2014).

\bibitem{dnc} D.N. Christodoulides, F. Lederer, and Y. Silberberg, Nature \textbf{424}, 817 (2003);

\bibitem{1Vortex} J. Cuevas, G. James, P.G. Kevrekidis, and K.H.J. Law, Physica D {\bf 238}, 1422 (2009).

\bibitem{yaron} H.S. Eisenberg, Y. Silberberg, R. Morandotti, A.R. Boyd, and J. S. Aitchison Phys. Rev. Lett. {\bf 81}, 3383 (1998).

\bibitem{yaron1} H.S. Eisenberg, Y. Silberberg, R. Morandotti, and J.S. Aitchison, Phys. Rev. Lett. {\bf 85}, 1863 (2000).

\bibitem{fetter2} A.L. Fetter, Rev. Mod. Phys. {\bf 81}, 647 (2009).

\bibitem{fetter1} A.L. Fetter, A.A. Svidzinsky, J. Phys.-Condens. Mat. {\bf 13}, R135 (2001).

\bibitem{fleischer} J.W. Fleischer, G. Bartal, O. Cohen, O. Manela, M. Segev, J. Hudock, and D. N. Christodoulides, Phys. Rev. Lett. {\bf 92}, 123904 (2004).

\bibitem{Victor3} M. Golubitsky, V.G. LeBlanc, I. Melbourne, J. Nonlinear Sci. {\bf 7}, 557 (1997).

\bibitem{christo2} R. Iwanow, D.A. May-Arrioja, D.N. Christodoulides, G.I. Stegeman, Y. Min, and W. Sohler, Phys. Rev. Lett. {\bf 95}, 053902 (2005).

\bibitem{johkiv} M. Johansson, and Yu.S. Kivshar, Phys. Rev. Lett. {\bf 82}, 85 (1999).

\bibitem{dnlsbook} P.G. Kevrekidis, {\it The Discrete Nonlinear Schr\"odinger Equation}, Springer-Verlag (Heidelberg, 2009).

\bibitem{siambook} P.G. Kevrekidis, D.J. Frantzeskakis, R. Carretero-Gonz{\'a}lez, {\it The defocusing nonlinear Schr{\"o}dinger equation: from dark solitons and vortices to vortex rings}, SIAM (Philadelphia, 2015).

\bibitem{sniper_yannis} P.G. Kevrekidis, I.G. Kevrekidis and A.R. Bishop, Phys. Lett. A {\bf 279}, 361 (2001).

\bibitem{hadii} P.G. Kevrekidis, H. Susanto, and Z. Chen, Phys. Rev. E 74, 066606 (2006).

\bibitem{Victor2} P. Kitanov, V.G. LeBlanc, SIAM J. Appl. Dyn. Syst. {\bf 16}, 16 (2017).

\bibitem{porto} H. Kim, G. Zhu, J.V. Porto, and M. Hafezi, Phys. Rev. Lett. {\bf 121}, 133002 (2018).

\bibitem{law} K.J.H. Law, H. Susanto, and P.G. Kevrekidis, Phys. Rev. A 78, 033802 (2008).

\bibitem{moti} F. Lederer, G.I. Stegeman, D.N. Christodoulides, G. Assanto, M. Segev, and Y. Silberberg, Phys. Rep. {\bf 463}, 1 (2008).

\bibitem{leykam}  D. Leykam and Y.D. Chong, Phys. Rev. Lett. {\bf 117}, 143901 (2016).

\bibitem{yaron2} R. Morandotti, U. Peschel, J.S. Aitchison, H.S. Eisenberg, and Y. Silberberg, Phys. Rev. Lett. {\bf 83}, 2726  (1999).

\bibitem{ober} O. Morsch and M. Oberthaler, Rev. Mod. Phys. {\bf 78}, 179 (2006).

\bibitem{neshev} D.N. Neshev, T.J. Alexander, E.A. Ostrovskaya, Yu. S. Kivshar, H. Martin, I. Makasyuk, and Z. Chen, Phys. Rev. Lett. {\bf 92}, 123903 (2004).

\bibitem{moti2} M.C. Rechtsman, J.M. Zeuner, Y. Plotnik, Y. Lumer, D. Podolsky, F. Dreisow, S. Nolte, M. Segev, and A. Szameit, Nature {\bf 496}, 196 (2013).

\bibitem{kip} C.E. R{\"u}ter, K.G. Makris, R. El-Ganainy, D.N. Christodoulides, M. Segev, and D. Kip, Nature Phys. {\bf 6}, 192 (2010).

\bibitem{SSW} B. Sandstede, A. Scheel, C. Wulff, J. Differ. Equations {\bf 141}, 122 (1997).

\bibitem{susjoh} H. Susanto and M. Johansson, Phys. Rev. E {\bf 72}, 016605 (2005).

\end{thebibliography}
\end{document}